\documentclass[aps,pre,twocolumn,10pt,superscriptaddress]{revtex4-2}
\usepackage{graphicx}
\usepackage{amsmath,units}
\usepackage{babel}
\usepackage{mathrsfs}
\usepackage{color} 
\usepackage[export]{adjustbox}
\allowdisplaybreaks[1]
\newcommand{\eref}[1]{Eq.~(\ref{#1})}%

\def\bea{\begin{eqnarray}}
\def\eea{\end{eqnarray}}

\def\nn{\nonumber}

\begin{document}
\title{Diffusion with Partial Resetting}
\author{Ofir Tal-Friedman}
\affiliation{School of Physics \& Astronomy, Raymond and Beverly Sackler Faculty of Exact Sciences, Tel Aviv University, Tel Aviv 6997801, Israel}

\author{Yael Roichman}
\affiliation{School of Physics \& Astronomy, Raymond and Beverly Sackler Faculty of Exact Sciences, Tel Aviv University, Tel Aviv 6997801, Israel}
\affiliation{School of Chemistry, Raymond and Beverly Sackler Faculty of Exact Sciences, Tel Aviv University, Tel Aviv 6997801, Israel}
\affiliation{Center for the Physics and Chemistry of Living Systems, Tel Aviv University, 6997801, Tel Aviv, Israel}

\author{Shlomi Reuveni}
\email{shlomire@tauex.tau.ac.il}
\affiliation{School of Chemistry, Raymond and Beverly Sackler Faculty of Exact Sciences, Tel Aviv University, Tel Aviv 6997801, Israel}
\affiliation{Center for the Physics and Chemistry of Living Systems, Tel Aviv University, 6997801, Tel Aviv, Israel}
\affiliation{The Sackler Center for Computational Molecular and Materials Science, Tel Aviv University, 6997801, Tel Aviv, Israel}

\begin{abstract}
Inspired by many examples in nature, stochastic resetting of random processes has been studied extensively in the past decade. In particular, various models of stochastic particle motion were considered where upon resetting the particle is returned to its initial position. Here we generalize the model of diffusion with resetting to account for situations where a particle is returned only a fraction of its distance to the origin, e.g., half way. We show that this model always attains a steady-state distribution which can be written as an infinite sum of independent, but not identical, Laplace random variables. As a result, we find that the steady-state transitions from the known Laplace form which is obtained in the limit of full resetting to a Gaussian form which is obtained close to the limit of no resetting. A similar transition is shown to be displayed by drift-diffusion whose steady-state can also be expressed as an infinite sum of independent random variables. Finally, we extend our analysis to capture the temporal evolution of drift-diffusion with partial resetting, providing a bottom-up probabilistic construction that yields a closed form solution for the time dependent  distribution of this process in Fourier-Laplace space. Possible extensions and applications of diffusion with partial resetting are discussed. 
\end{abstract}

\maketitle

\section{Introduction}
\begin{figure}[t]
	\centering
	\includegraphics[width=0.45\textwidth]{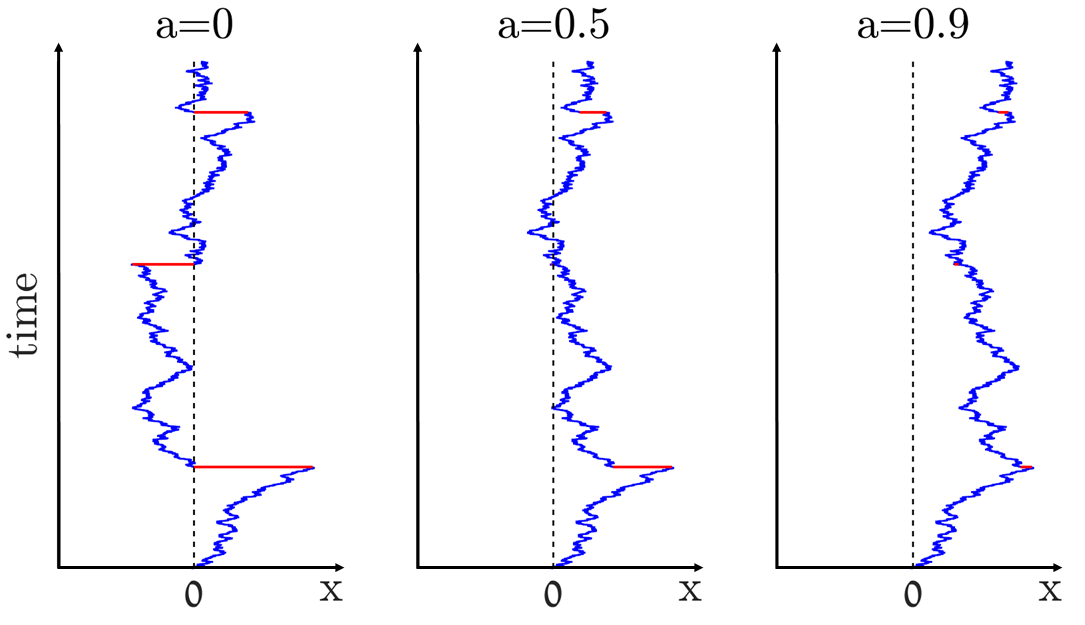}
	\caption{Diffusion with partial resetting. A particle undergoing diffusion is stochastically reset to a position $x'= ax$, with rate $r$. The left panel, where $a=0$, represents the classical Evans-Majumdar model of diffusion with (full) stochastic resetting \cite{evans_diffusion_2011}. The center and right panels show diffusion with partial resetting where $a=0.5$ and $a=0.9$, respectively.}
	\label{fig:abstract}
\end{figure}

Random motion under resetting has been studied extensively in the past decade both theoretically  \cite{eliazar_searching_2007,kusmierz_first_2014,kusmierz_optimal_2015,roldan_stochastic_2016,nagar_diffusion_2016,reuveni_optimal_2016,pal_first_2017,falcao_interacting_2017,montero_continuous-time_2017,shkilev_continuous-time_2017,maes_induced_2017,evans_run_2018,chechkin_random_2018,pal_first_2019,pal_local_2019,santos_fractional_2019,ahmad_first_2019,ray_peclet_2019,bodrova_nonrenewal_2019,maso-puigdellosas_transport_2019,pal_search_2020,ray_diffusion_2020,riascos_random_2020,bressloff_modeling_2020,pinsky_diffusive_2020,bressloff_directed_2020,de_bruyne_optimization_2020,bodrova_continuous-time_2020,gonzalez_diffusive_2021,singh_resetting_2020,stojkoski2022income,vinod2022nonergodicity,PhysRevE.104.024105} and recently, also experimentally \cite{tal-friedman_experimental_2020,besga_optimal_2020,faisant_optimal_2021}. It has been established that an unbound random motion becomes asymptotically bound once resetting to the origin is initiated, thus leading to a new type of non-equilibrium steady-state \cite{evans_diffusion_2011,evans_diffusion_2014,pal_diffusion_2015,pal_diffusion_2016,mendez_characterization_2016,montero_directed_2016,eule_non-equilibrium_2016,pal_invariants_2019,masoliver_telegraphic_2019,bodrova_scaled_2019,pal_time-dependent_2019,kusmierz_subdiffusive_2019,gupta_stochastic_2019,bodrova_resetting_2020,bodrova_brownian_2020,evans_stochastic_2020,miron_diffusion_2021,PhysRevE.104.014121}. For example, the probability to find a colloidal particle diffusing in a suspending fluid at position $x$ at time $t$ is given by the known Gaussian form, $P(x,t)=\frac{1}{\sqrt{4\pi D t}}e^{-x^2/4Dt}$, where $D$ is the diffusion constant. In this case, we have normal diffusion and the mean squared displacement of the particle diverges linearly with time. In contrast, if the particle is returned to the origin stochastically with rate $r$ (Fig. \ref{fig:abstract}, left panel), it will become confined to the vicinity of the origin such that at long times its position distribution will converge to a steady-state that is given by the Laplace distribution: $P(x)=\frac{\alpha_0}{2}e^{-\alpha_0|x|}$, where $\alpha_0=\sqrt{r/D}$ is an inverse length scale corresponding to the typical distance diffused by the particle in the time between two consecutive resetting events \cite{evans_diffusion_2011}.

Full resetting amounts to a situation where the value of a given observable is initialized to zero (or any other value), thus erasing all memory of past events. While this extreme form of resetting is the one most widely studied to date, one can easily imagine situations where resetting acts in a partial manner, e.g. when a catastrophic event leads to partial extinction of a growing population \cite{gripenberg1983stationary,ben2019random}. As another example, consider a case where resetting acts to backtrack a diffusing particle to one of its previously visited locations according to some law, as was e.g. done in  \cite{boyer_solvable_2014,boyer_long_2017,falcon-cortes_localization_2017,santos_non-gaussian_2018,campos_recurrence_2019}. It is then natural to ask if any type of `backtrack resetting' will result in a stationary position distribution. For example, will a drift-diffusion process that has a directed motion component arrive at a steady-state even for infinitesimally weak backtracking? 

Here, we study this and related questions via diffusion with \textit{partial resetting} which acts to return the particle part of its way back to the origin. An example is given in the middle and left panels of Fig.~\ref{fig:abstract}, where a diffusing  particle is partially reset at stochastic times from its position $x$ to a new position $x'=ax$, with $0\leq a < 1$. We note that a similar model in which $a$ was considered random was analyzed and solved for ballistic motion in \cite{dahlenburg_stochastic_2021}. Here, we go beyond pure ballistic motion and show that partial resetting leads to a highly non-trivial, yet fully tractable, time-dependent and steady-state behaviour for diffusion with and without drift.

We note that steady-state and first-passage properties of the model considered herein were  studied in parallel and independently by J. Kevin Pierce in a paper that appeared on the arXiv while we were writing this manuscript \cite{pierce_advection-diffusion_2022}.
Previously,  the same author obtained the steady-state distribution for the model in chapter 5 of his Ph.D. thesis \cite{pierce2021stochastic}, where he considered a stochastic description of bedload sediment transport. This was done using methods different from the ones employed below, was unknown to us, and brought to our attention in a personal communication only after our manuscript appeared on the arXiv.


The remainder of this paper is structured as follows. In Sec. \ref{Diffusion with Partial Resetting}, we introduce the model of diffusion with partial resetting and show that it always leads to a bound steady-state position distribution to which we provide a closed-form expression. We moreover show that this steady-state distribution transitions from the known Laplace form which is obtained in the limit of full resetting ($a=0$) \cite{evans_diffusion_2011} to a Gaussian form which is obtained close to the limit of no resetting ($a=1$). In Sec. \ref{Drift Diffusion}, we go on to study drift-diffusion with partial resetting. Here too, we show that---despite drift being present---partial resetting leads to a confined steady-state distribution for any value of the partial resetting parameter ($0\leq a<1$). We provide a close-form expression for this steady-state distribution, and show that it too transitions from a known form which is obtained in the limit of full resetting \cite{pal_diffusion_2015,pal_invariants_2019} to a Gaussian form which is obtained close to the limit of no resetting. In Sec. \ref{Sharp Spatial Restart}, we analyze a deterministic partial resetting protocol in which resetting occurs at fixed time intervals, i.e., sharp resetting \cite{bhat_stochastic_2016,pal_diffusion_2016,pal_first_2017,eliazar_branching_2017,eliazar_mean-performance_2020}. We analyze the time evolution of this process and show that it converges to a cyclo-stationary steady-state. The insight gained from the analysis of the sharp resetting mechanism is carried over to Sec. \ref{Time Dependent Solution} in which we build the full time-dependent probability distribution of drift-diffusion with stochastic partial resetting, from the bottom up. We conclude in Sec. \ref{Conclusions}, where we discuss possible extensions and applications of diffusion with partial resetting.

\section{Diffusion with Partial Resetting} \label{Diffusion with Partial Resetting}
Consider diffusion in the presence of partial stochastic resetting. A particle starts its motion at the origin and diffuses until resetting occurs. The resetting process is stochastic: times between consecutive resetting events are taken from an exponential distribution with rate  $r$. When resetting occurs the particle's position undergoes an instantaneous transformation 
\begin{equation} 
\label{eqn:par_res}
x\xrightarrow[]{\substack{partial \\ resetting}}ax~, 
\end{equation}
with $0\leq a \leq 1$. Thus, when $a=0$, the particle is brought back to its initial position, and in the other extreme limit, when $a=1$, no resetting occurs and the particle continues diffusing unaffected. For intermediate values of $a$, partial resetting occurs: the particle is taken to an intermediate position in between its final position and the origin.

The master equation describing diffusion with partial resetting is given by
\begin{equation} 
\label{eqn:spatialexp}
\frac{\partial P(x,t)}{\partial t}=D\frac{\partial^2 P(x,t)}{\partial x^2}-rP(x,t)+\frac{r}{a} P(x/a,t)~,
\end{equation}
where $P(x,t)$ is the probability to find the particle at position $x$ at time $t$, $D$ is the diffusion constant, and $r$ is the resetting rate. The change of probability density, at position $x$ and time $t$, has three contributions. The first term on the right hand side accounts for diffusion, the second term accounts for probability loss at $x$ due to resetting with rate $r$, and the third term accounts for probability gain at $x$ due to resetting at $x/a$ with rate $r$. Note that in this latter case the probability flow into the small interval $[x,x+\delta]$ comes from partial resetting occurring at the small interval $[x/a,x/a+\delta/a]$. This interval is larger by a factor of $1/a$, which explains why the resetting rate in the third term is scaled by the same amount. 

At the steady state Eq. \eqref{eqn:spatialexp} reduces to 
\begin{equation}
\label{eqn:fokker_ss}
    D\frac{d^2 P(x)}{d x^2}-rP(x)+\frac{r}{a} P(x/a)=0~,
\end{equation}
which we Fourier transform to obtain
\begin{equation}
\label{eqn:Fm}
    -(r+Dk^2)\hat{P}(k)+r\hat{P}(ak)=0~.
\end{equation}
The solution to Eq. \eqref{eqn:Fm} can be shown to be given by 
\begin{equation}
\label{eqn:Fss}
     \hat{P}_{ss}(k)= \prod_{j=0}^{\infty}\frac{r}{r+Dk^2 a^{2j}}~,
\end{equation}
which is verified in Appendix A. 

The result in Eq. \eqref{eqn:Fss} extends the result derived by Evans and Majumdar for diffusion with (full) stochastic resetting. Indeed, taking $a=0$, we have $\hat{P}_{ss}(k)= {r}/({r+Dk^2})$ which can be inverted to give $P_{ss}(x)=\frac{\alpha_0}{2} exp(-\alpha_0|x|)$ with $\alpha_0=\sqrt{r/D}$ as found in \cite{evans_diffusion_2011}. More generally, for $0<a<1$, the product form of Eq. \eqref{eqn:Fss} implies that the steady state position of the particle $X_{ss}$, admits the following stochastic representation
\begin{equation}
\label{eqn:sum}
X_{ss}=\sum_{j=0}^{\infty}X_j~,    
\end{equation}
where \{$X_0,X_1,X_2,...$\} are independent Laplace random variables. 

To see this, we recall that the Fourier transform of a Laplace distribution with variance $\sigma^2$ and density  $f(x)=\frac{1}{\sqrt{2}\sigma}exp(-\sqrt{2}|x|/ \sigma)$ is given by
\begin{equation}
    \hat{f}(k)=\frac{1}{1+\sigma^2k^2/2}~.
    \label{laplace_fourier}
\end{equation}
Comparing with Eq.~\eqref{eqn:Fss}, and
due to the fact that the Fourier transform of a 
sum of independent random variables is the product of their Fourier transforms, we see that each $X_j$ in Eq. \eqref{eqn:sum} is a Laplace random variable with variance 
\begin{equation}
\sigma_j^2=2Da^{2j}/r~.
\label{variance_laplace}
\end{equation}
We thus conclude that the steady-state position distribution of diffusion with partial stochastic resetting can be expressed as an infinite sum of independent, but not identical, Laplace random variables. Note, that all these random variables have zero mean and that their variance drops exponentially with the running index $j$, thus making their contribution to the sum in Eq. \eqref{eqn:sum} smaller and smaller.    

\begin{figure}[t]
    \centering
   	\includegraphics[width=0.425 \textwidth]{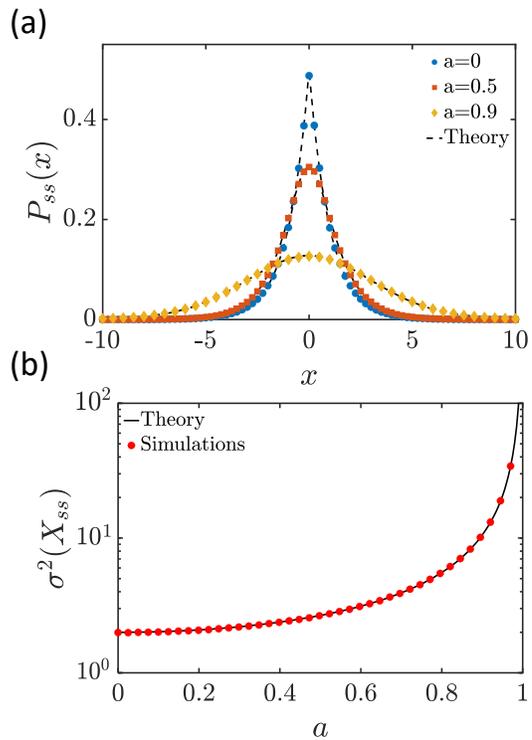}
    \caption{Steady-state and variance of diffusion with partial stochastic resetting. Panel (a): steady-state probability distributions for different values of the partial resetting parameter $0\leq a<1$. Very good agreement is seen between simulations for $a=0, 0.5, 0.9$ (ordered top to bottom, marked by circles, squares, and diamonds respectively), and direct sampling from the theoretical steady-state distribution of Eq. \eqref{eqn:sum} (dashed lines).  Panel (b): The variance of the steady-state distribution as a function of the partial resetting parameter $0\leq a<1$. The theoretical prediction of Eq. \eqref{eqn:var_eq} is plotted as a solid black curve and red circles depict results coming from simulations. Throughout this figure, the diffusion constant and resetting rate were set to $D=1/100$ and $r=1/100$ respectively.}
    \label{fig:hist+var}
\end{figure}
    
In Fig. 2a, we plot the solution for different values of the partial resetting parameter $0\leq a<1$. We do this by sampling directly from the infinite sum of Laplace distributions that is presented in Eq. \eqref{eqn:sum}, where here and in what follows we approximate this sum by its first 100 terms. This result is compared with direct numerical simulations of diffusion with partial stochastic resetting. It can be seen that the steady-state position distribution is centered around the origin. This is clear by symmetry, and also by the fact that all the random variables on the right hand side of \eref{eqn:sum} have zero mean. Thus, the first moment of the steady-state distribution vanishes identically. 

We also observe that the steady-state distribution becomes wider as $a\xrightarrow{}1$, i.e., in the limit of weak partial resetting. Indeed, utilizing the independence of the random variables in \eref{eqn:sum}, we find that the variance of the steady-state position distribution is given by 
\begin{equation}
\label{eqn:var_eq}
    \sigma^2(X_{ss})=\sum_{j=0}^\infty \sigma_j^2=\dfrac{2D}{r}\dfrac{1}{1-a^2}~,
\end{equation}
which diverges at $a=1$ as expected for free diffusion without resetting (Fig. \ref{fig:hist+var}b). Yet, note that a steady-state of finite variance is attained whenever $a < 1$.

\begin{figure}[t!]
\centering
	\includegraphics[width=0.4\textwidth]{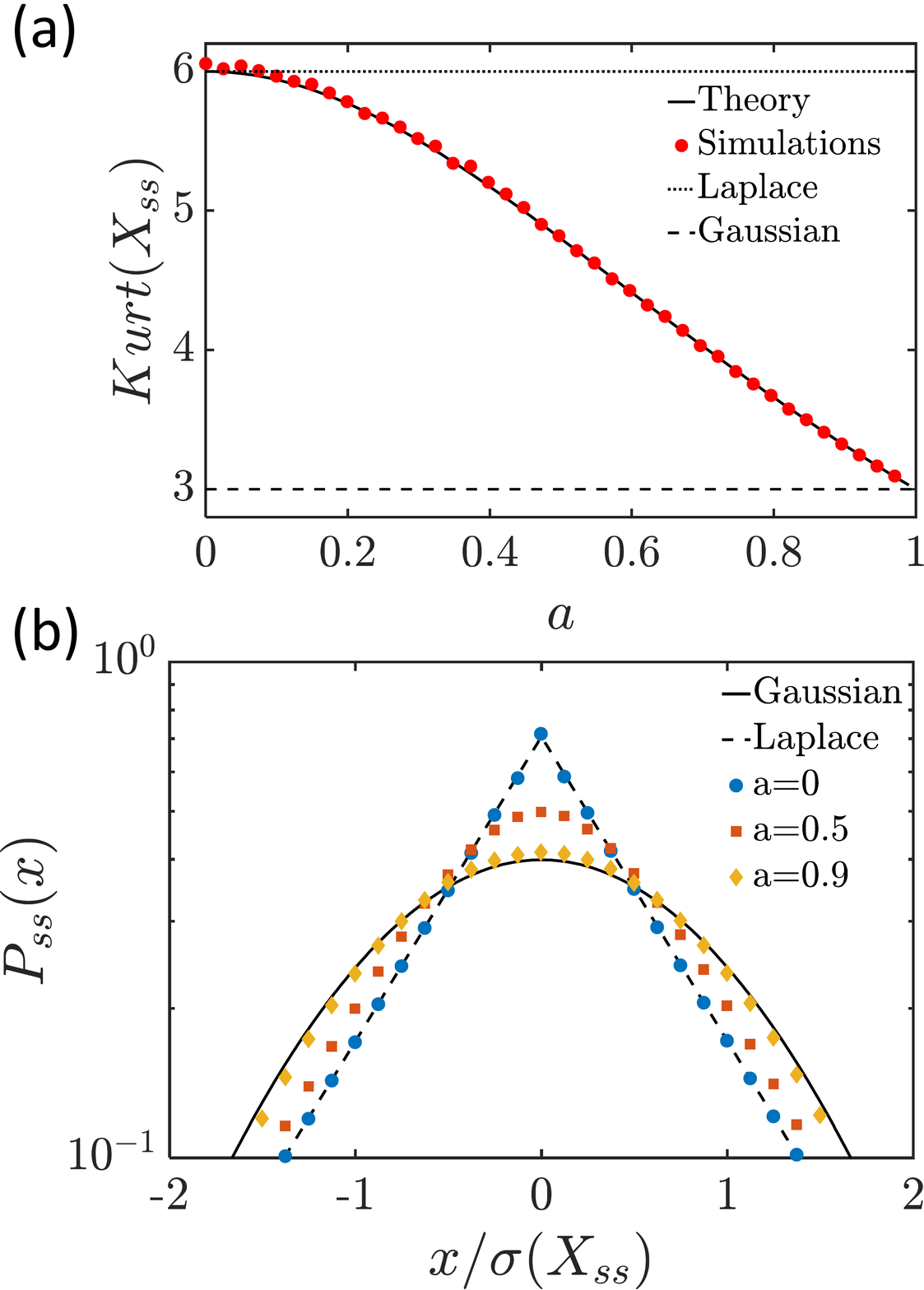}
    \caption{The steady-state distribution of diffusion with partial resetting transitions from Laplace to Gaussian. Panel (a): the kurtosis from \eref{eqn:kurtosis} is plotted as a function of the partial resetting parameter $0\leq a <1$. Very good agreement is seen between the theoretical prediction (solid black line) and results coming from simulations (red circles). Recall that the Laplace and Gaussian distributions have kurtosis six and three correspondingly, which is indicated by horizontal dashed  lines. Panel (b): The steady state probability distributions from Fig. 2a normalized by their standard deviations. A clear transition from the Laplace law to the Gaussian law is observed as $a$ is tuned from zero to unity.}\label{fig:log+kurt}
\end{figure}

While the third moment of the steady-state position distribution also vanishes by symmetry, the fourth moment does not. To compute it, we observe that higher moments of the random variables appearing on the right hand side of Eq. \eqref{eqn:sum} can be computed directly from their distribution, which combined with their independence gives 
\begin{equation}
\label{fourthmom}
    \langle X^4_{ss} \rangle  = \dfrac{12D^2}{r^2}\dfrac{1}{1-a^4}+ 3\left( \dfrac{2D}{r}\dfrac{1}{1-a^2}\right)^2~, \end{equation}
as we show in Appendix B. Combining Eqs. \eqref{eqn:var_eq} and (\ref{fourthmom}), we obtain the kurtosis 
\begin{equation}
     \text{Kurt}(X_{ss})= \dfrac{\langle X^4_{ss} \rangle}{\text{Var}(X_{ss})^2}=\dfrac{6}{1+a^2}~,\
     \label{eqn:kurtosis}
\end{equation}
which transitions from the value of six to the value of three as the partial resetting parameter $a$ is tuned in the range $[0,1]$ (Fig. \ref{fig:log+kurt}a). 

The kurtosis and its dependence on the partial resetting parameter implies that the steady-state distribution transitions from the Laplace distribution which is obtained in the limit of full resetting ($a=0$) to a nearly Gaussian distribution that is obtained close to the limit of no resetting ($a=1$). This means that the shape of the steady-state distribution can be controlled by tuning the value of $a$. The transition between the Laplace and Gaussian forms is illustrated by scaling the steady-state distributions from Fig.~\ref{fig:hist+var}a by their standard deviations and plotting them in Fig. \ref{fig:log+kurt}b. 

We have already seen that the Laplace distribution emerges in the limit of full resetting. To also see the Gaussian limit analytically, consider the distribution of $X_{ss}/\sigma(X_{ss})$, i.e., the steady-state position scaled by its standard deviation. The Fourier transform of this random variable follows from Eq. \eqref{eqn:Fss} and is given by 
\begin{equation}
\label{eqn:Fss_scaled1}
     \langle e^{-ikX_{ss}/\sigma(X_{ss})} \rangle= \prod_{j=0}^{\infty}\frac{1}{1+\frac{Dk^2 a^{2j}}{r \sigma^2(X_{ss})}}~.
\end{equation}
Fixing $k$ and taking the limit $a\rightarrow1$, we have $Dk^2/r \sigma^2(X_{ss})\ll1$, which yields the following approximation 
\begin{equation}
\label{eqn:Fss_scaled2}
     \langle e^{-ikX_{ss}/\sigma(X_{ss})} \rangle \simeq \prod_{j=0}^{\infty}e^{-\frac{Dk^2 a^{2j}}{r \sigma^2(X_{ss})}}=e^{-k^2/2}~.
\end{equation}
This proves the result as the right hand side is nothing but the Fourier transform of a Gaussian random variable with zero mean and unit variance. 

\section{Drift-Diffusion with Partial Resetting} \label{Drift Diffusion}

We now turn our attention to the case of drift-diffusion with partial resetting. Under full resetting, it is well established that drift-diffusion attains a steady-state \cite{pal_diffusion_2015,pal_invariants_2019}. However, here partial resetting can be made arbitrarily weak by taking the limit $a \rightarrow 1$. As drift and partial resetting compete, one may have expected that in this limit resetting would be too weak to confine the particle. Yet, we will now show that a confined steady-state always emerges.

We start with the master equation, which for drift-diffusion with partial resetting reads
\bea
\label{eqn:fokker_plank_non_ss}
\frac{\partial P(x,t)}{\partial t}&=&D\frac{\partial^2 P(x,t)}{\partial x^2}-V\frac{\partial P(x,t)}{\partial x} \nn \\
&&-rP(x,t)+\frac{r}{a} P(x/a,t)~,
\eea
where $V$ is the drift velocity. At the steady state this equation reduces to
\begin{equation}
\label{eqn:fokker_v}
    D\frac{d^2 P(x)}{d x^2}-V\frac{d P(x)}{d x}-rP(x)+\frac{r}{a} P(x/a)=0~,
\end{equation}
which we Fourier transform to obtain
\begin{equation}
\label{eqn:Fm_V}
    -(r+Dk^2-ikv)\hat{P}(k)+r\hat{P}(ak)=0~.
\end{equation}
The solution to \eref{eqn:Fm_V} is given by 
\begin{equation}
\label{eqn:Fss_V}
     \hat{P}_{ss}(k)= \prod_{j=0}^{\infty}\frac{r}{r-iVka^j+Dk^2 a^{2j}}~, 
\end{equation}
which is verified in Appendix C.

The result in Eq. \eqref{eqn:Fss_V} extends the known result for drift-diffusion with (full) stochastic resetting \cite{pal_diffusion_2015,pal_invariants_2019}. Indeed, taking the limit $a\to0$, we have $\hat{P}_{ss}(k)= {r}/({r-iVk+Dk^2})$ which can be inverted to give $P_{ss}(x)= \frac{\alpha_0}{2\sqrt{1+\lambda_0^2}} e^{-(\sqrt{1+\lambda_0^2}-sgn(x)\lambda_0)\alpha_0|x|}$, with $\alpha_0=\sqrt{r/D}$ and $\lambda_0 = V/(2\sqrt{Dr})$ (Appendix D). More generally, for $0<a<1$, the product form of Eq. \eqref{eqn:Fss_V} implies that the steady state position of the particle $X_{ss}$, admits the following stochastic representation
\begin{equation}
\label{eqn:sumV}
X_{ss}=\sum_{j=0}^{\infty}X_j,    
\end{equation}
where \{$X_0,X_1,X_2,...$\} are independent random variables coming from the same family.

To see this, observe that Eq. \eqref{eqn:Fss_V} asserts that the Fourier transform of $X_j$ in Eq. \eqref{eqn:sumV} is given by
\begin{equation}
    \hat{P_j}(k) = \langle e^{-ikX_j} \rangle={r}/({r-iV_jk+D_jk^2}),
    \label{laplace_fourier_V}
\end{equation}
where $V_j=Va^j$ and $D_j=Da^{2j}$. We thus conclude that the steady-state position distribution of drift-diffusion with partial stochastic resetting can be expressed as an infinite sum of independent, but not identical, random variables whose densities are given by \begin{equation}
\label{eqn:arnab_drift}
P_{j}(x) = \frac{\alpha_j}{2\sqrt{1+\lambda^2_j}} e^{-(\sqrt{1+\lambda^2_j}-sgn(x)\lambda_j)\alpha_j|x|}, 
\end{equation}
where $\alpha_j = \sqrt{r/D_j}$ and $\lambda_j = V_j/(2\sqrt{D_jr})$. 

\begin{figure}[t!]
    \centering
	\includegraphics[width=0.45\textwidth]{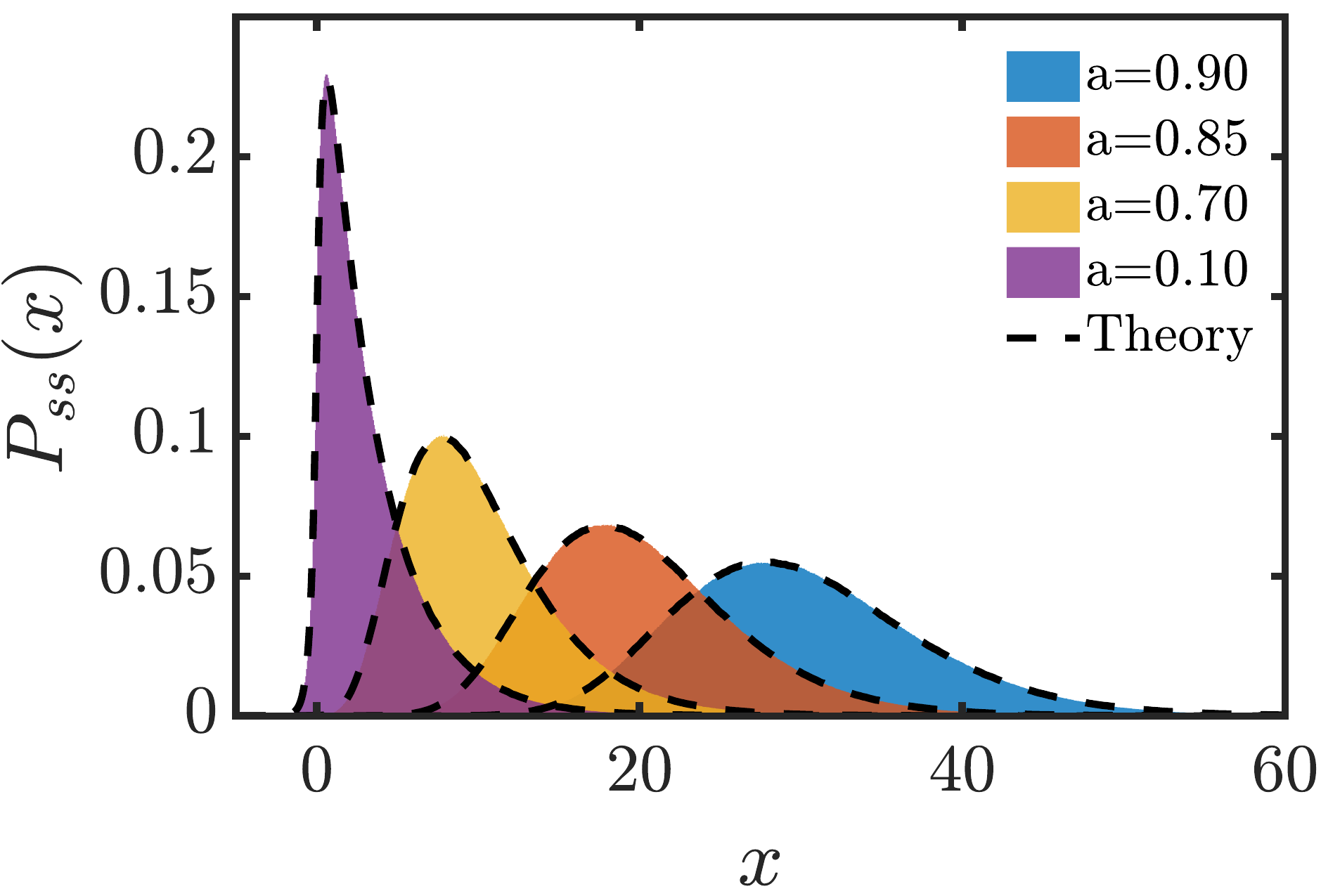}
    \caption{Steady-state probability distributions for drift-diffusion with different values of the partial resetting parameter $a=0.1, 0.7, 0.85,0.9$, which appear from left to right in ascending order of mean position. Very good agreement is seen between simulations (colored histograms) and direct sampling from the theoretical steady-state distribution of Eq. \eqref{eqn:sumV} (dashed lines). Here, the drift velocity, diffusion constant, and resetting rate were set to $V=3/100$, $D=1/100$ and $r=1/100$ respectively.}
    \label{fig:0.1}
\end{figure}

In Fig. \ref{fig:0.1}, we plot the solution for different values of the partial resetting parameter $0\leq a<1$. We do this by sampling directly from the infinite sum presented in Eq. \eqref{eqn:sumV}, which is compared with direct numerical simulations of drift-diffusion with partial stochastic resetting. It can be seen that both the mean and variance of the steady-state position distribution grow as $a\xrightarrow{}1$. Indeed, taking expectations in \eref{eqn:sumV}, we find that the mean of the steady-state distribution is given by 
\begin{equation}
\label{mean_V}
    \langle X_{ss} \rangle  = \sum_{j=0}^{\infty}  \langle X_{j} \rangle=\sum_{j=0}^{\infty} V_j/r=\frac{V}{r}\frac{1}{1-a}~. 
\end{equation}
Similarly, utilizing the independence of the random variables in \eref{eqn:sumV}, we find 
\bea
  \label{var_V}
  \sigma^2(X_{ss})&=&\sum_{j=0}^\infty \sigma_j^2 
  =\sum_{j=0}^\infty \left(\frac{2D_j}{r} +\frac{V_j^2}{r^2}\right) \nn \\
&=& \left(\dfrac{2D}{r}+\dfrac{V^2}{r^2}\right)\dfrac{1}{1-a^2}~. 
\eea
\noindent Thus, we see that while the mean and variance both diverge in the limit $a \rightarrow 1$, they remain finite even when partial resetting is very weak, and as long as $a<1$. We also note that by taking the special case of pure drift $D=0$, we obtain the same expressions for the steady-state mean and variance as those that are predicted by Eqs. (68) and (69) of ref. \cite{dahlenburg_stochastic_2021}, when the limit of a fixed (deterministic) resetting amplitude is taken there.

We now show that in the limit $a \rightarrow 1$ the steady-state distribution is approximately Gaussian. To see this, consider the distribution of $\left( X_{ss} - \langle X_{ss} \rangle \right)/\sigma(X_{ss})$, i.e., the standardized position. The Fourier transform of this random variable is given by 
\bea
\label{DD_pre_approx}
  \langle e^{-\frac{ik\left( X_{ss} - \langle X_{ss} \rangle \right)}{\sigma(X_{ss})}} \rangle&=& \prod_{j=0}^{\infty}\langle e^{-\frac{ik\left( X_j - \langle X_j \rangle \right)}{\sigma(X_{ss})}} \rangle~,
\eea
where we have used the stochastic representation of Eq. \eqref{eqn:sumV} and the independence of the random variables there. Fixing $k$ and taking the limit $a\rightarrow1$, we have $k/\sigma(X_{ss})\ll1$. Expanding the exponents on the right hand side of Eq. \eqref{DD_pre_approx} to second order and taking expectations yields the following approximation 
\begin{equation}
    \langle e^{-\frac{ik\left( X_j - \langle X_j \rangle \right)}{\sigma(X_{ss})}} \rangle \simeq 1- \frac{k^2\sigma_j^2}{2\sigma^2(X_{ss})} \simeq e^{-\frac{k^2\sigma_j^2}{2\sigma^2(X_{ss})}}~.
\end{equation}
It follows that in this limit 
\begin{equation}
\label{eqn:Fss_scaled_V2}
    \langle e^{-\frac{ik\left( X_{ss} - \langle X_{ss} \rangle \right)}{\sigma(X_{ss})}} \rangle \simeq e^{-k^2/2}~,
\end{equation}
which once again proves the result since the right hand side of \eref{eqn:Fss_scaled_V2} is nothing but the Fourier transform of a Gaussian random variable with zero mean and unit variance. 

\section{Sharp Partial Resetting}\label{Sharp Spatial Restart}
So far, we have assumed that resetting is conducted stochastically with rate $r$. Another interesting case to consider is that of resetting at constant time intervals of duration $\tau$. To tackle this common form of resetting \cite{bhat_stochastic_2016,pal_diffusion_2016,pal_first_2017,eliazar_branching_2017,eliazar_mean-performance_2020}, also known as sharp resetting, we will present a probabilistic argument that circumvents the need to solve the corresponding (generalized) master equation. The insight gained from this approach to the solution will also prove useful in the next section where we will present a bottom-up construction of the  time-dependent probability distribution of drift-diffusion with partial resetting at a constant rate $r$. 

To this end, we once again start by writing the spatial distribution of a particle which diffused freely for a time $t$ that is smaller than the sharp resetting time $\tau$, i.e., $t<\tau$. This is simply given by the known Gaussian form
\begin{equation}
\label{eqn:gauss}
    P(x,t)=\frac{1}{\sqrt{4\pi D t}}e^{-\frac{x^2}{4Dt}}~.
\end{equation}
At $t=\tau$ partial resetting occurs, taking the particle from its random position $x$ to $ax$. Since the particle's position at the resetting moment comes from a Gaussian distribution with density given by Eq. \eqref{eqn:gauss}, the particle's position immediately after resetting, i.e., at time $t\rightarrow\tau^+$, is also Gaussian with density
\begin{equation}
\label{eqn:gauss_2}
    P(x,\tau^+)=\frac{1}{\sqrt{4\pi D a^2 \tau}}e^{-\frac{x^2}{4Da^2\tau}}~,
\end{equation}
which is obtained by scaling a Gaussian random variable by the partial resetting parameter $a$. Note, that this is also the probability distribution that describes a particle that diffused freely for time $a^2\tau$. Thus, the combined effect of free diffusion for time $\tau$ and consecutive partial resetting with parameter $a$ is equivalent to the effect of free diffusion for an effective time $\tau_{\text{eff}}=a^2\tau$.

\begin{figure}[t]
	\centering
	\includegraphics[width=0.45\textwidth]{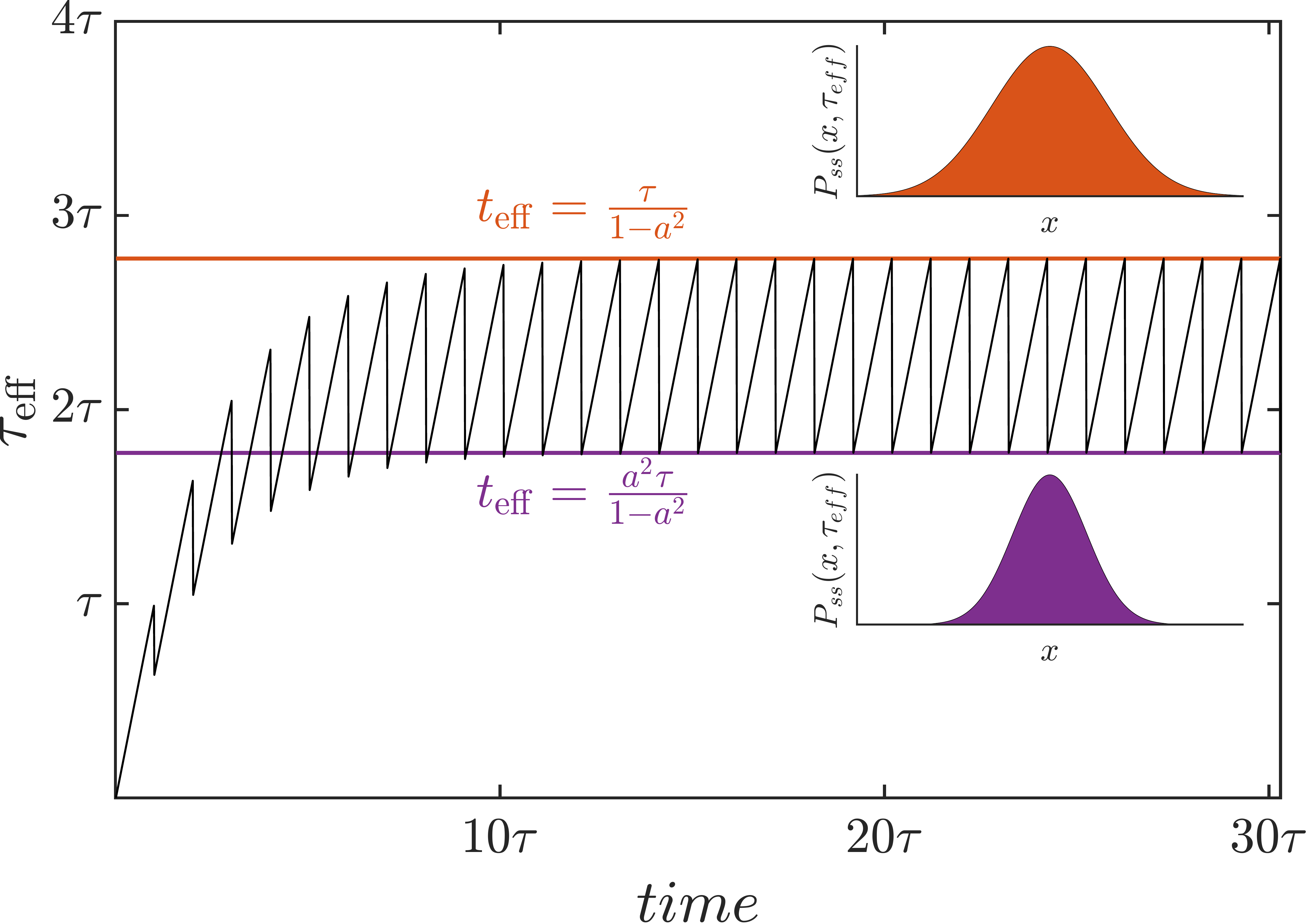}
	\caption{The temporal behaviour of diffusion with  sharp partial resetting is governed by an effective diffusion time $\tau_{\text{eff}}$ whose own time evolution is in turn governed by the process described by \eref{eqn:tau_sharp}. At long times, the effective diffusion time cycles between a low value and and high value as shown in the figure. Immediately after resetting occurs, the position distribution is Gaussian with an effective diffusion time  $\tau_{\text{eff}} = \frac{a^2\tau}{1-a^2}$. Resetting then takes exactly $\tau$ units of time to occur. Thus, right before resetting occurs, the position distribution is Gaussian with an effective diffusion time of $\tau_{\text{eff}} = \frac{\tau}{1-a^2}$. In this figure, the diffusion constant and the partial resetting parameter were set to $D=1$ and $a=0.8$ respectively.}
	\label{fig:sharp}
\end{figure}

As sharp resetting is conducted at fixed time intervals, this process now repeats itself periodically. Namely, free diffusion for time $\tau$ is followed by partial resetting with parameter $a$, and so on and so forth. Each diffusion period adds $\tau$ to the effective diffusion time, and each resetting event multiplies the effective diffusion time by a factor of $a^2$. Thus, for the effective diffusion time we have
\begin{equation}
\label{eqn:tau_sharp}
     0 \xrightarrow[]{\mathcal{D}}\tau\xrightarrow[]{\mathcal{R}}  a^2\tau\xrightarrow[]{\mathcal{D}} a^2\tau+\tau \xrightarrow[]{\mathcal{R}} a^2(a^2\tau +\tau)\dots~,
\end{equation}
where $\mathcal{D}$ stands for a diffusion period and $\mathcal{R}$ stands for a partial resetting event. Asymptotically this process converges to a Gaussian cyclic steady state whose probability density is given by
\begin{equation}
\label{eqn:cyclic_steady_state}
    P(x,\tau_{\text{eff}})=\frac{1}{\sqrt{4\pi D \tau_{\text{eff}}}}e^{-\frac{x^2}{4D\tau_{\text{eff}}}}~,
\end{equation}
where the effective diffusion time oscillates between a high value $\tau_{\text{eff}} = \sum_{n=0}^{\infty} a^{2n}\tau=\frac{\tau}{1-a^2}$ which is attained at the end of every diffusion period and a low value $\tau_{\text{eff}} =\sum_{n=1}^{\infty} a^{2n}\tau= \frac{a^2\tau}{1-a^2} $ which is attained immediately after a resetting event occurred. This cycle and the corresponding Gaussian distributions at both its ends, are illustrated in Fig. \ref{fig:sharp}.

\section{Time Dependent Solution} 
\label{Time Dependent Solution}

In Secs. \ref{Diffusion with Partial Resetting} and \ref{Drift Diffusion} we have dealt with the steady-state of diffusion with partial stochastic resetting. We will now go on to consider the temporal evolution of this process. Instead of going for a brute-force solution of Eqs. \eqref{eqn:spatialexp} and \eqref{eqn:fokker_plank_non_ss}, we will offer a probabilistic analysis from which insight can be drawn. This analysis will be based on the results established in the previous section. 

We start with a diffusing particle which is also subject to partial stochastic resetting that is conducted at a constant rate $r$. Assuming that there were exactly $m$ resetting events in the time interval $[0,t]$, we denote their occurrence times by $\{t_1,...,t_m\}$. Using the same arguments presented in Sec. \ref{Sharp Spatial Restart}, we obtain an effective diffusion time of 
\begin{equation}
\label{eqn:T_m_def}
    \tau_{\text{eff}}(m)=t-a^{2m}[1-a^2]\sum_{n=1}^m t_n a^{-2n}~,
\end{equation}
which is obtained by applying the process in \eref{eqn:tau_sharp} with $t_1$ as the diffusion time till the first resetting event, $t_2-t_1$ as the diffusion time between the first and second resetting events, and so on in a similar manner until the observation time $t$ is reached. A detailed derivation of Eq. \eqref{eqn:T_m_def} is given in Appendix E. 

It follows that the probability distribution to find the particle at position $x$ at time $t$---given that exactly $m$ resetting events occurred at times $\{t_1,...,t_m\}$---can be written as

\begin{equation}
\label{eqn:P_m_def}
     P(x,t,\tau_{\text{eff}}(m))= \frac{1}{\sqrt{4\pi D \tau_{\text{eff}}(m)}} e^{-\frac{x^2}{4D\tau_{\text{eff}}(m)}}~.
\end{equation}
To obtain the unconditional probability distribution of the position at time $t$, the above result must be averaged with proper weights over all possible resetting time epochs $\{t_1,...,t_m\}$ and over all the possible numbers of resetting events.

Since resetting is conducted at a constant rate $r$, the probability to have $m$ resetting events in the time interval $[0,t]$ is given by the Poisson distribution
 \begin{equation}
 \label{eqn:pois}
     Pr(m,t) = \frac{(rt)^me^{-rt}}{m!}.
 \end{equation}
Averaging Eq. \eqref{eqn:P_m_def} over $m$ using Eq. \eqref{eqn:pois} gives
\begin{equation}
     \sum_{m=0}^{\infty} \frac{(rt)^me^{-rt}}{m!\sqrt{4\pi D \tau_{\text{eff}}(m)}} e^{-\frac{x^2}{4D\tau_{\text{eff}}(m)}},
\end{equation}
which can be Fourier transformed to give
\begin{equation}
  \sum_{m=0}^{\infty} \hat{P}_m(k,t,\tau_{\text{eff}}(m))\equiv\sum_{m=0}^{\infty}\frac{(rt)^m}{m!}e^{-rt-Dk^2\tau_{\text{eff}}(m)}~.
\end{equation}

The effective time $\tau_{\text{eff}}(m)$ is defined in Eq. \eqref{eqn:T_m_def} as a function of the observation time $t$ and the resetting time epochs $\{t_1,...,t_m\}$. We now recall that the basic properties of the Poisson process assert that if $m$ resetting events occurred their statistics is uniform in the range $[0,t]$. We thus have 
\begin{align}
\label{eqn:time_avg_p_hat}
    &\hat{P}_m(k,t)= \nn \\ \nn \\ 
    &\int_0^t dt_m \int_0^{t_m} dt_{m-1}  \cdots\int_0^{t_{2}}\frac{m!}{t^m}\hat{P}_m(k,t,\tau_{\text{eff}}(m)) dt_1~,
\end{align}
and note that the $m!$ accounts for equally likely permutations which were lost by assuming  $t_1<t_2<...<t_m<t$ in the integrals above.

Calculating the integrals in Eq. \eqref{eqn:time_avg_p_hat}, and Laplace transforming $\tilde{P}_m(k,s)=\int_0^{\infty}\hat{P}_m(k,t)e^{-st}dt$, we find (Appendix F)
\begin{align}
\label{Fourier-Laplace_pre_solution}
&\Tilde{P}_0(k,s)=\frac{1}{r+Dk^2+s}~,\nn\\
&\Tilde{P}_1(k,s)=\frac{r}{(r+Dk^2a^2+s)}\Tilde{P}_0(k,s)~,\nn\\
&\Tilde{P}_2(k,s)=\frac{r}{(r+Dk^2a^4+s)}\Tilde{P}_1(k,s)~, 
\end{align}
from which we deduce the time dependent probability distribution in Fourier-Laplace space 
\begin{equation}
\label{Fourier-Laplace_solution}
    \Tilde{P}(k,s)=\sum_{m=0}^\infty\Tilde{P}_m(k,s)=\sum_{m=0}^\infty\frac{1}{r}\prod_{j=0}^m\frac{r}{ r+Dk^2a^{2j}+s}~.
\end{equation}

\noindent The steady-state solution of Eq. \eqref{eqn:Fss} can be obtained by taking the long time limit of the above results. Formally, this is done by using the Final Value Theorem, $\lim_{t \to \infty} f(t) =\lim_{s \to 0} s\Tilde{f}(s)$, see Appendix G for details.

\begin{figure}[t]
	\centering
	\includegraphics[width=0.45\textwidth]{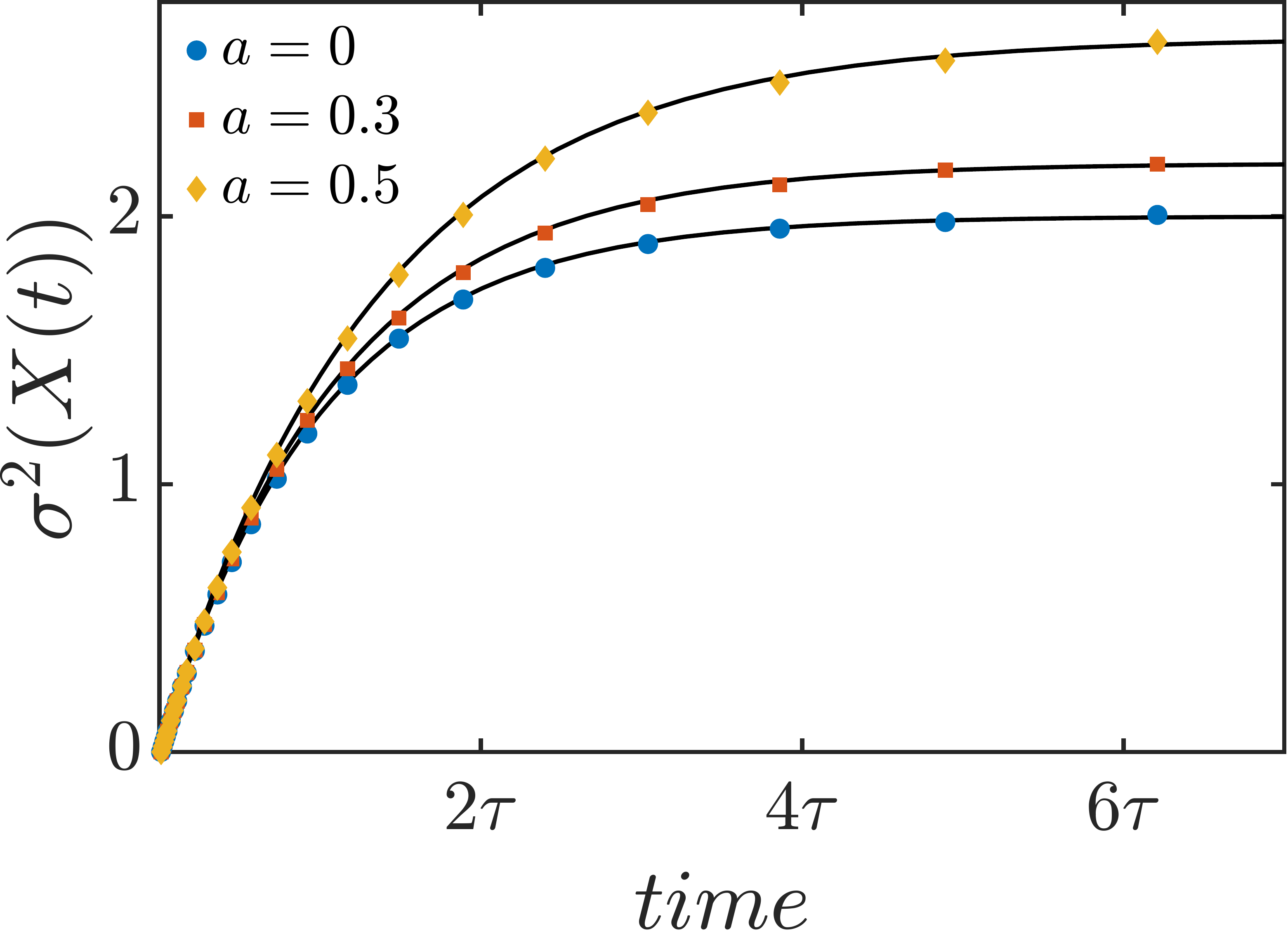}
	\caption{The time dependent variance of  diffusion with partial resetting for different values of the partial resetting parameter $a$. Very good agreement is seen between simulations for $a=0, 0.3, 0.5$ (marked by circles, squares, and diamonds respectively), and the theoretical result of Eq. \eqref{eqn:var_of_time_depen} (solid lines).}
	\label{fig:td_var}
\end{figure}

While the mean of the time dependent distribution vanishes, Eq. \eqref{Fourier-Laplace_solution} can be used to obtain the time dependent variance (Appendix H)
\begin{equation}
\label{eqn:var_of_time_depen}
    \sigma^2(X(t)) = \frac{2D}{r(1-a^2)}\left(1-e^{-r(1-a^2)t}\right),
\end{equation}
which is plotted in Fig. \ref{fig:td_var}. This variance converges exponentially to the steady-state variance of Eq. \eqref{eqn:var_eq}. Interestingly, the rate of convergence is proportional to $1-a^2$, which asserts that convergence times will be long in the limit $a\to1$.

We note that the products appearing in Eq. \eqref{Fourier-Laplace_solution} are similar to the product that appears in Eq. \eqref{eqn:Fss} for the steady-state. There are, however, two differences: (i) an additional $s$ in the denominator of each term that appears in the products of Eq. \eqref{Fourier-Laplace_solution}; and (ii) the products of Eq.  \eqref{Fourier-Laplace_solution} being finite rather than infinite. Following this observation, one can guess the time dependent solution of drift-diffusion with partial resetting from its steady-state that was found in Eq. \eqref{eqn:Fss_V}. This reads
\begin{equation}
\label{eqn:drift_difussion_pd}
    \Tilde{P}(k,s)=\sum_{m=0}^\infty\frac{1}{r}\prod_{j=0}^m\frac{r}{ r-iVka^j+Dk^2a^{2j}+s}~.
\end{equation}
\noindent and it can be readily verified that this probability distribution solves
\begin{equation}
\label{eqn:fourier_laplace_rearranged}
    \Tilde{P}(k,s)=\frac{r\Tilde{P}(ak,s)+1}{r-iVk+Dk^2+s}~,
\end{equation}
which is the rearranged Fourier-Laplace transform of Eq. \eqref{eqn:fokker_plank_non_ss}. This is shown in Appendix I.

\section{Conclusions} \label{Conclusions}
In this paper, we extended the Evans-Majumdar model of diffusion with stochastic resetting \cite{evans_diffusion_2011}. Rather than full resetting, we considered a case where the particle is returned only part way back to the origin, such that upon resetting $x \rightarrow ax$ with $0\leq a<1$. We found that this resetting protocol always results in a steady-state whose Fourier transform we gave in closed form. This, in turn, allowed us to show that the steady-state distribution can be understood as an infinite sum of independent Laplace random variables with increasingly smaller variance. Moreover, we showed that the steady-state distribution interpolates smoothly between the Laplace steady-state distribution, which is found for full resetting ($a=0$), and a Gaussian distribution which is obtained in the limit $a\rightarrow 1$. 

We then extended our analysis to drift-diffusion with partial resetting where we have shown that a steady-state emerges even when partial resetting is very weak. The latter can once again be expressed as a sum of independent random variables with increasingly smaller variance. Similar to the no-drift case, the steady-state of drift-diffusion with partial resetting also undergoes a transition between a non-Gaussian steady-state with exponential tails and a Gaussian steady-state, as the partial resetting parameter $a$ is tuned from zero to unity. An interesting application of this result is the possibility to mimic the effect of different confining potentials with partial resetting, for example in an optical tweezers setup as was used in \cite{tal-friedman_experimental_2020}. Here we obtained probability densities which are equivalent to those of a particle diffusing in an harmonic and linear potentials, as well as a non-trivial interpolation between these densities. Other effective potentials may be obtained by considering more sophisticated versions of partial resetting, e.g., via resetting kernels which are discussed below. 

Having established the steady-state properties of diffusion with partial resetting, we turned to investigate its time evolution. To do so, we first considered a sister problem: diffusion with sharp partial resetting, i.e., partial resetting that is conducted periodically at constant time intervals. We analyzed the time evolution of this process and showed that it leads to a cyclo-stationary steady-state. The insight gained from this analysis was then carried over to the original problem where resetting is conducted stochastically at a constant rate. In particular, rather than solving the time-dependent master equation directly, we presented a probabilistic analysis which allowed us to construct the solution step-by-step from the bottom up. In this way, we were able to obtain an analytical closed-form expression for the Fourier-Laplace transform of the time dependent probability distribution describing diffusion with partial resetting with, and without, drift.

The model considered in this paper falls into a broader class of models which can be described by the following master equation
\begin{eqnarray} 
\label{eqn:expand_x}
\frac{\partial P(x,t)}{\partial t} &=& D\frac{\partial^2 P(x,t)}{\partial x^2} -V\frac{\partial P(x,t)}{\partial x} \\ \nonumber
&-& rP(x,t) + r \int_{-\infty}^{\infty}K(x',x)P(x',t) dx'~.
\end{eqnarray}
Here, the top row describes diffusion with drift, and the second row the effect of resetting at a constant rate $r$. Specifically, the first term in the second row accounts for the probability loss due to resetting at $x$, and the second term the probability gain due to resetting transition that take the particle back to $x$ from other locations. The kernel $K(x',x)$ describes the rules of the game by defining a probability distribution over all possible positions $x$ given the position $x'$. In this paper we focused on a specific resetting kernel, $K(x',x)=\delta(x'-\frac{x}{a})/a$. A different choice of resetting kernel may lead to fundamentally different results e.g., the absence of a steady-state distribution. Questions such as this have been studied in a similar system, where the resetting kernel was taken to be time-dependent rather than space-dependent    \cite{boyer_solvable_2014,boyer_long_2017,falcon-cortes_localization_2017,santos_non-gaussian_2018,campos_recurrence_2019}. It was shown that motion under these conditions is bound only if the memory kernel decays fast enough with time, else the MSD of the particle diverges. In the future, it would be interesting to consider the general case where the resetting kernel depends both on time and space.\\

\begin{acknowledgments}
Shlomi Reuveni acknowledges support from the Israel Science Foundation (grant No. 394/19). This project has received funding from the European Research Council (ERC) under the European Union’s Horizon 2020 research and innovation programme (Grant agreement No. 947731). Yael Roichman acknowledges support from the Israel Science Foundation (grants No. 988/17 and 385/21).
\end{acknowledgments}

\newpage

\onecolumngrid
\appendix

\appendix
\section{Corroboration of Eq. \eqref{eqn:Fss}}
At the steady-state, the master equation boils down to  
\begin{equation}
    D \frac{d^2 P(x)}{d x^2} -rP(x)+\frac{r}{a} P(x/a)=0~.
\end{equation}
\noindent Recalling some Fourier transform identities

\begin{center}
    $\mathcal{F}\left(\frac{df(x)}{dx}\right)=ik \hat{f}(k) \ ; \ \mathcal{F}\left(f(\frac{x}{a})\right)=a\hat{f}(ak)~,$
\end{center}
we obtain
\begin{equation}
\label{eqn:Fm_appendix}
    -(r+Dk^2)\hat{P}(k)+r\hat{P}(ak)=0~.
\end{equation}

\noindent Substituting Eq. \eqref{eqn:Fss} into the left hand side of Eq. \eqref{eqn:Fm_appendix}, we find 
\bea
\text{L.H.S.}&=&-(r+Dk^2)\prod_{j=0}^{\infty}\frac{r}{r+Dk^2\cdot a^{2j}} 
+r\cdot\prod_{j=0}^{\infty}\frac{r}{r+Dk^2a^2\cdot a^{2j}} \nn \\
&=& \prod_{j=1}^{\infty}\frac{r}{r+Dk^2\cdot a^{2j}} \left(\frac{-r(r+Dk^2)}{r+Dk^2}+r\right) 
=0~,
\eea
which proves that Eq. \eqref{eqn:Fss} is indeed a steady-state solution of diffusion with partial resetting. 

\section{Moments of the steady-state distribution, variance, and kurtosis}
The moments of the steady-state distribution of diffusion with partial resetting can be found by using the following relation
\begin{equation}
    \label{eqn:momentseq}
    \langle X_{ss}^m \rangle = (-i)^m\frac{d^m\hat{P}_{ss}(k)}{dk^m}\bigg| _{k=0}~,
\end{equation}
where $\hat{P}_{ss}(k)$ is given by Eq. \eqref{eqn:Fss}. In Eq. \eqref{eqn:sum} we have shown that $X_{ss}$ can be written as an infinite sum of independent Laplace random variables \{$X_0,X_1,X_2,...$\}, such that 
\begin{equation}
    \langle e^{-ikX_j} \rangle \equiv \hat{P_j}(k)=\frac{r}{r+Dk^2 a^{2j}}~.
\end{equation}
\noindent Taking derivatives, we find
\begin{gather}
       \label{eqn:derivaties}
    \frac{d\hat{P_j}(k)}{dk}= -\dfrac{2a^{2j}Drk}{\left(a^{2j}Dk^2+r\right)^2} \\
    \frac{d^2\hat{P_j}(k)}{dk^2}=\dfrac{2a^{2j}Dr\left(3a^{2j}Dk^2-r\right)}{\left(a^{2j}Dk^2+r\right)^3} \\
    \frac{d^3\hat{P_j}(k)}{dk^3}=-\dfrac{24a^{4j}D^2rk\left(a^{2j}Dk^2-r\right)}{\left(a^{2j}Dk^2+r\right)^4} \\
    \frac{d^4\hat{P_j}(k)}{dk^4}=\dfrac{24a^{4j}D^2r\left(5a^{4j}D^2k^4-10a^{2j}Drk^2+r^2\right)}{\left(a^{2j}Dk^2+r\right)^5},
\end{gather}
\noindent resulting in the following moments:
\begin{gather}
\label{eqn:moments}
    \langle X_j \rangle = 0~, \\
    \langle X_j^2 \rangle = \dfrac{2Da^{2j}}{r} = \sigma^2(X_{j})~, \\
    \langle X_j^3 \rangle = 0~, \\
    \langle X_j^4 \rangle = \dfrac{24D^2a^{4j}}{r^2}~. 
\end{gather}

Using the fact that the variance of the sum of independent random variables is equal to the sum of their variances, we can immediately find the variance of the steady-state position. This is given by
\begin{equation}
    \sigma^2(X_{ss})=\sum_{j=0}^\infty \dfrac{2D}{r}a^{2j}=\dfrac{2D}{r}\dfrac{1}{1-a^2}~.
\end{equation}
To find the fourth moment, we observe that
\begin{gather*}
    \langle X_{ss}^4 \rangle =\langle (X_0+X_1+...)^4 \rangle =  \sum_{i=0}^{\infty} \langle X_i^4 \rangle + 3\sum_{i=0}^\infty \sum_{j=0}^\infty \langle X^2_i \rangle \langle X^2_j \rangle - 3\sum_{i=0}^\infty \langle X_i^2\rangle^2 \\
    =\dfrac{24D^2}{r^2}\dfrac{1}{1-a^4}+ 3\sum_{i=0}^\infty \langle X^2_i \rangle  \sum_{j=0}^\infty \langle X^2_j \rangle -\dfrac{12D^2}{r^2}\dfrac{1}{1-a^4}\\
    = \dfrac{12D^2}{r^2}\dfrac{1}{1-a^4}+ 3\left( \dfrac{2D}{r}\dfrac{1}{1-a^2}\right)^2,
    \end{gather*}
where we have once again used independence and the fact that moments of odd order vanish to either simplify or kill mixed terms. Given the fourth moment, the kurtosis follows from its definition
\begin{gather*}
     \text{Kurt}(X_{ss})= \dfrac{\langle X_{ss}^4 \rangle}{\sigma^4(X_{ss})}=\dfrac{\dfrac{12D^2}{r^2}\dfrac{1}{1-a^4}+ 3\left( \dfrac{2D}{r}\dfrac{1}{1-a^2}\right)^2}{\left(\dfrac{2D}{r}\dfrac{1}{1-a^2}\right)^2} 
     =\dfrac{6}{1+a^2}~.
\end{gather*}

\section{Corroboration of Eq. \eqref{eqn:Fss_V}}
At the steady-state, the master equation boils down to  
\begin{equation}
    D \frac{d^2 P(x)}{d x^2} - V\frac{dP(x)}{dx}-rP(x)+\frac{r}{a} P(x/a) = 0~.
\end{equation}
\noindent Similarly to Appendix A, we Fourier transform and obtain
\begin{equation}
\label{eqn:Fm_appendix_v}
    -(r+Dk^2-ikV)\hat{P}(k)+r\hat{P}(ak) = 0~.
\end{equation}
\noindent Substituting Eq. \eqref{eqn:Fss_V} into the left hand side of Eq. \eqref{eqn:Fm_appendix_v}, we find
\bea
\text{L.H.S.}&=&-(r-iVk+Dk^2)\prod_{j=0}^{\infty}\frac{r}{r-iVka^j+Dk^2 a^{2j}} 
+r\cdot\prod_{j=0}^{\infty}\frac{r}{r-iVka\cdot a^j+Dk^2a^2\cdot a^{2j}} \nn \\
&=& \prod_{j=1}^{\infty}\frac{r}{r+Dk^2\cdot a^{2j}-iVka^j} \left(-r+r\right)=0~,
\eea
\noindent which proves that Eq. \eqref{eqn:Fss_V} is indeed a steady-state solution of drift-diffusion with partial resetting.


\section{Steady-State of drift-diffusion under (full) stochastic resetting}

Here we simply recall the result for drift-diffusion under (full) stochastic   resetting which can e.g. be found in ref. \cite{pal_invariants_2019}. The probability density reads
\begin{equation}
\label{eqn:arnab_eq}
\frac{\sqrt{\frac{r}{D}}}{2\sqrt{1+\frac{V^2}{4Dr}}}exp \left[-\sqrt{\frac{r}{D}}|x|\left(\sqrt{1+\frac{V^2}{4Dr}}- \frac{V\cdot sgn(x)}{2\sqrt{Dr}}\right)\right]~,
\end{equation}
which simplifies to  
\begin{equation}
   P_{ss}(x)=\frac{1}{\sqrt{\frac{4D}{r}+\frac{V^2}{r^2}}}e^{-\sqrt{\frac{r}{D}+\frac{V^2}{4D^2}}|x|+\frac{V}{2D}x}~.
\end{equation}
This expression is of the shape of $c\cdot exp(-a|x|+bx)$, which has a known Fourier transform \cite{oberhettinger_fourier_2014} 
\begin{equation}
    \mathcal{F} \left(c\cdot exp(-a|x|+bx)\right)=\frac{2ac}{a^2-(b+ik)^2}~.
\end{equation}

\noindent Substituting $a=\sqrt{\frac{r}{D}+\frac{V^2}{4D^2}}$, $b=\frac{V}{2D}$, and $c=\frac{1}{\sqrt{\frac{4D}{r}+\frac{V^2}{r^2}}}$ we obtain 
\bea
P_{ss}(x)&=&2\frac{\sqrt{\frac{r}{D}+\frac{V^2}{4D^2}}}{\sqrt{\frac{4D}{r}+\frac{V^2}{r^2}}}\frac{1}{\frac{r}{D}+\frac{V^2}{4D^2}-(\frac{V}{2D}+ik)^2} \nn \\
&=& \frac{r}{r-ikV+Dk^2}~,
\eea
which is the same as Eq. \eqref{eqn:Fss_V} for $a=0$.

\section{Derivation of Eq. \eqref{eqn:T_m_def}}
Similarly to Sec. \ref{Sharp Spatial Restart}, we will describe the process as diffusion with an effective diffusion time $\tau_{\text{eff}}$. However, note that the time intervals between partial resetting events are no longer constant. To obtain the  effective diffusion time at time $t$, we recall that resetting occurred at times $\{t_1,...,t_m\}$, which are all smaller than $t$. The time intervals between resetting events are thus of lengths $\{t_1,t_2-t_1,...,t_m-t_{m-1}\}$, and one must not forget that free diffusion also occurs at the last time interval  $(t_{m},t]$ which comes after the final resetting event. It follows that the effective diffusion time evolves according to     
\begin{gather}
\label{eqn:tau_depend}
     0 \xrightarrow[]{\mathcal{D}}t_1\xrightarrow[]{\mathcal{R}}  a^2 t_1 \xrightarrow[]{\mathcal{D}} a^2 t_1 +t_2-t_1 \nn \\
     \xrightarrow[]{\mathcal{R}} a^2(a^2 t_1 +t_2-t_1) \nn \\ \xrightarrow[]{\mathcal{D}} a^2(a^2 t_1 +t_2-t_1)+t_3-t_2~,...
\end{gather}
These terms can be rearranged to give 
\begin{gather}
\label{eqn:tau_depend_arr}
     0 \xrightarrow[]{\mathcal{D}}t_1\xrightarrow[]{\mathcal{R}}  a^2 t_1 \xrightarrow[]{\mathcal{D}} (a^2-1) t_1 + t_2 \nn\\
     \xrightarrow[]{\mathcal{R}} (a^4-a^2) t_1 + a^2t_2
     \xrightarrow[]{\mathcal{D}} (a^4-a^2) t_1 + (a^2-1)t_2 + t_3\nn\\
     \dots, \xrightarrow[]{\mathcal{D}} t+\sum_{n=1}^m (a^{2m+2-2n}-a^{2m-2n})t_n~,
\end{gather}
which in turn yields the following effective diffusion time 
\begin{gather}
    \tau_{\text{eff}}(m) = t+\sum_{n=1}^m (a^{2m+2-2n}-a^{2m-2n})t_n = t-a^{2m}[1-a^2]\sum_{n=1}^m t_n a^{-2n}~,
\end{gather}
that appears in Eq. \eqref{eqn:T_m_def}.
\onecolumngrid
\section{Derivation of Eq. \eqref{Fourier-Laplace_pre_solution}}
We start by calculating $\hat{P}_m(k,t)$ which we will Laplace transform in the second stage. For $m=0$, the effective diffusion time is equal to the observation time. Thus,  $\tau_{\text{eff}}(m)=t$, which gives 
\begin{equation}
    \hat{P}_0(k,t)=e^{-(r+Dk^2)t}~.
\end{equation}
For $m\geq1$ we follow Eq. \eqref{eqn:time_avg_p_hat}. For example, for $m=1$ we obtain 
\begin{align}
\hat{P}_1(k,t)&=\int_0^t\frac{1}{t}rte^{-rt}e^{-Dk^2(t+(a^2-1)t_1)}dt_1 \nn \\
              &=r\int_0^t e^{-rt}e^{-Dk^2(t+(a^2-1)t_1)}dt_1 \nn \\
              &=re^{-(r+Dk^2)t}\int_0^te^{-Dk^2(a^2-1)t_1}dt_1 \nn \\
              &=\frac{-re^{-(r+Dk^2)t}}{Dk^2[a^2-1]}\left( e^{-Dk^2(a^2-1)t} -1 \right)~.
\end{align}
Similarly, for $m=2$ we obtain 
\begin{align}
\hat{P}_2(k,t)&=\int_0^tdt_2\int_0^{t_2}\frac{(rt)^2e^{-rt}}{t^2}e^{-Dk^2(t+(a^2-1)(t_1 a^{2} +t_2))}dt_1 \nn \\
&=r^2e^{-(r+Dk^2)t}\int_0^tdt_2\int_0^{t_2}e^{-Dk^2(a^2-1)(t_1a^2+t_2)}dt_1 \nn \\
&=r^2e^{-(r+Dk^2)t}\int_0^tdt_2\left[\frac{e^{-Dk^2(a^2-1)(t_1a^2+t_2)}}{-Dk^2(a^2-1)a^2}\right]_{t_1=0}^{t_2} \nn \\
&=\frac{-r^2e^{-(r+Dk^2)t}}{Dk^2(a^4-a^2)} \int_0^tdt_2\left[e^{-Dk^2(a^4-1)t_2} - e^{-Dk^2(a^2-1)t_2}\right] \nn \\
&=\frac{r^2e^{-(r+Dk^2)t}}{(Dk^2)^2(a^4-a^2)} \left[\frac{e^{-Dk^2(a^4-1)t}-1}{a^4-1}  -\frac{e^{-Dk^2(a^2-1)t}-1}{a^2-1}\right]~.
\end{align}

Next, we Laplace transform the above expressions to find 
\begin{equation}
    \tilde{P}_0(k,s)=\frac{1}{r+Dk^2+s}~,
\end{equation}
and 
\begin{align}
\Tilde{P}_1(k,s)&=\frac{-r}{Dk^2[a^2-1]}\left[ \frac{1}{r+Dk^2+(a^2-1)Dk^2+s}-\frac{1}{r+Dk^2+s}\right] \nn \\
&=\frac{-r}{Dk^2[a^2-1]}\left[ \frac{1}{r+a^2Dk^2+s}-\frac{1}{r+Dk^2+s}\right] \nn \\
&=\frac{-r}{Dk^2[a^2-1]}\frac{(1-a^2)Dk^2}{(r+a^2Dk^2+s)(r+Dk^2+s)} \nn \\
&=\frac{r}{(r+a^2Dk^2+s)(r+Dk^2+s)}~.
\end{align}
Similarly, for $m=2$ we find
\begin{align}
\Tilde{P}_2(k,s)&=\frac{r^2}{(Dk^2)^2(a^2-1)a^2}\left[\frac{1}{a^4-1}\left(\frac{1}{r+Dk^2a^4+s}-\frac{1}{r+Dk^2+s} \right) \right. \nn \\
&\left.-\frac{1}{a^2-1}\left(\frac{1}{r+Dk^2a^2+s}-\frac{1}{r+Dk^2+s} \right)\right]\nn \\
&=\frac{r^2}{(Dk^2)^2(a^2-1)a^2}\left[\frac{1}{a^4-1}\frac{(1-a^4)Dk^2}{(r+Dk^2a^4+s)(r+Dk^2+s)} \right. \nn \\
&\left.-\frac{1}{a^2-1}\frac{(1-a^2)Dk^2}{(r+Dk^2a^2+s)(r+Dk^2+s)}  \right] \nn \\
&=\frac{r^2}{Dk^2(a^2-1)a^2}\left[\frac{1}{(r+Dk^2a^2+s)(r+Dk^2+s)}-\frac{1}{(r+Dk^2a^4+s)(r+Dk^2+s)}\right]\nn \\
&=\frac{r^2}{Dk^2(a^2-1)a^2}\frac{Dk^2(a^4-a^2)}{(r+Dk^2+s)(r+Dk^2a^2+s)(r+Dk^2a^4+s)}\nn \\
&=\frac{r^2}{(r+Dk^2+s)(r+Dk^2a^2+s)(r+Dk^2a^4+s)}
\end{align}
\section{Recovering the steady-state distribution from the time-dependent solution}
We use the Final Value Theorem,
\begin{equation}
    \lim_{t \to \infty} f(t) =\lim_{s \to 0} s\Tilde{f}(s),
\end{equation}
where $\Tilde{f}(s)$ is the Laplace transform of $f(t)$. In our case, we have
\begin{equation}
    \hat{P}_{ss}(k) = \lim_{s \to 0} s\Tilde{P}(k,s).
\end{equation}
Plugging in the time dependent distribution of diffusion with partial resetting from Eq. \eqref{Fourier-Laplace_solution}, we obtain
\begin{equation}
\lim_{s \to 0} s\Tilde{P}(k,s)=\lim_{s \to 0}\sum_{m=0}^\infty\frac{s}{r}\prod_{j=0}^m\frac{r}{r+Dk^2a^{2j}+s}.
\end{equation}
Defining $N=r/s$, we rewrite the above expression as
\begin{equation}
\lim_{N \to \infty}\frac{1}{N}\sum_{m=0}^N\prod_{j=0}^m\frac{r}{r+Dk^2a^{2j}+s} \equiv \lim_{N \to \infty}\frac{1}{N}\sum_{m=0}^NF(N,m),
\end{equation}
where
\begin{equation}
F(N,m) = \prod_{j=0}^m\frac{r}{r+Dk^2a^{2j}+r/N}.
\end{equation}
We now observe that 
\begin{equation}
\lim_{N \to \infty} F(N,m) = \prod_{j=0}^m\frac{r}{r+Dk^2a^{2j}} \equiv F(m) ,
\end{equation}
and that 
\begin{equation}
\lim_{m \to \infty} F(m) = \prod_{j=0}^\infty\frac{r}{r+Dk^2a^{2j}} = \hat{P}_{ss}(k) ,
\end{equation}
where in the last step we simply recalled the definition of the steady-state distribution in \eref{eqn:Fss}. Now, since the  average of the first $n$ elements in an infinite series that converges to a limit also converges to the same limit, we have  
\begin{equation}
\lim_{N \to \infty}\frac{1}{N}\sum_{m=0}^N\prod_{j=0}^m\frac{r}{r+Dk^2a^{2j}+s} \equiv \lim_{N \to \infty}\frac{1}{N}\sum_{m=0}^NF(N,m)=\hat{P}_{ss}(k).
\end{equation}
This proves that the steady-state solution of Eq. \eqref{eqn:Fss} can be obtained by taking the long time limit of Eq. \eqref{Fourier-Laplace_solution}.
\section{Time dependent variance}
In Laplace space, the second moment of the time-dependent probability distribution is given by:
\begin{equation}
    \langle X^2(s) \rangle = -\frac{d^2\Tilde{P}(k,s)}{dk^2}\bigg| _{k=0}.~
\end{equation}
We define
\begin{equation}
    f(j) = \frac{r}{r+Dk^2a^{2j}+s},
\end{equation}
and write the time dependent solution for diffusion with partial resetting as
\begin{equation}
    \Tilde{P}(k,s)=\sum_{m=0}^\infty\Tilde{P}_m(k,s)=\frac{1}{r}\sum_{m=0}^\infty\prod_{j=0}^mf(j)~.   
\end{equation}
The first derivative of $\Tilde{P}(k,s)$ can be written as
\begin{equation}
    \frac{d\Tilde{P}(k,s)}{dk}=\frac{1}{r}\sum_{m=0}^\infty\left(\left( \prod_{j=0}^{m}f(j)\right)\left( \sum_{j=0}^{m}\frac{f'(j)}{f(j)} \right) \right),
\end{equation}
where $f'(j) \equiv df(j)/dk$.
The second derivative of $\Tilde{P}(k,s)$ can be written as
\begin{equation}
\label{eqn:sec_deriv_td}
    \frac{d^2\Tilde{P}(k,s)}{dk^2}=\frac{1}{r}\sum_{m=0}^\infty\left(\left( \prod_{j=0}^{m}f(j)\right)\left( \sum_{j=0}^{m}\frac{f'(j)}{f(j)}\right)^2 +\left( \prod_{j=0}^{m}f(j)\right)\left( \sum_{j=0}^{m}\frac{f''(j)\cdot f(j) - f'(j)^2}{f(j)^2}\right)\right).
\end{equation}
The first and second derivative of each $f(j)$ is given by 
\bea
    f'(j) &=& - \frac{2rDka^{2j}}{\left( r+Dk^2a^{2j}+s\right)^2}, \nn \\
    f''(j) &=& -\frac{2rDa^{2j}}{\left( r+Dk^2a^{2j}+s\right)^2}+\frac{2rDa^{2j}k\left(4(r+s)Da^{2j}k+4D^2a^{4j}k^3\right)}{\left( r+Dk^2a^{2j}+s\right)^4},
\eea
and after substituting $k=0$,
\bea
f'(j)\big| _{k=0} &=& 0, \nn \\
f''(j)\big| _{k=0} &=& -\frac{2rDa^{2j}}{\left( r+s \right)^2}.
\eea
Substituting these derivatives into Eq. \eqref{eqn:sec_deriv_td} results in
\bea
    \frac{d^2\Tilde{P}(k,s)}{dk^2}\bigg| _{k=0}&=&\frac{1}{r}\sum_{m=0}^\infty \left(\prod_{j=0}^{m}\left(\frac{r}{r+s}\right)\right)\left( \sum_{j=0}^m -\frac{2rDa^{2j}}{\left( r+s \right)^2}\cdot\frac{r+s}{r}  \right) \nn \\
    &=&-2D\sum_{m=0}^\infty\left(\frac{r}{r+s}\right)^{m+1}\sum_{j=0}^m\frac{a^{2j}}{r(r+s)} \nn \\
&=&-\frac{2D}{(r+s)^2}\sum_{m=0}^\infty\left(\frac{r}{r+s}\right)^{m}\frac{1-a^{2(m+1)}}{1-a^2} \nn \\
&=&-\frac{2D}{(r+s)^2} \frac{1}{1-a^2}\sum_{m=0}^\infty\left(\frac{r}{r+s}\right)^{m}-a^2\left(\frac{ra^2}{r+s}\right)^{m} \nn \\
&=&-\frac{2D}{(r+s)^2} \frac{1}{1-a^2}\left(\frac{1}{1-\frac{r}{r+s}} -\frac{a^2}{1-\frac{ra^2}{r+s}}\right) \nn \\
&=&-\frac{2D}{(r+s)^2} \frac{1}{1-a^2}\left( \frac{r+s}{s} - \frac{a^2(r+s)}{(1-a^2)r+s}\right) \nn \\
&=&-\frac{2D}{1-a^2}\left(\frac{1}{s(r+s)}-a^2\frac{1}{(r+s)(r(1-a^2)+s)}\right),
\eea
which we invert to find the time-dependent variance of diffusion with partial resetting
\begin{equation}
    \sigma^2(X)=\langle X^2(t) \rangle = \mathscr{L}^{-1}\left(-\frac{d^2\Tilde{P}(k,s)}{dk^2}\bigg| _{k=0}\right) = \frac{2D}{r(1-a^2)}\left(1-e^{-r(1-a^2)t}\right),
\end{equation}
which is \eref{eqn:var_of_time_depen} in the main text. Note that for $t\rightarrow\infty$ this expression reduces to the variance found for the steady-state of diffusion with partial resetting (\eref{eqn:var_eq} in the main text). 

\section{Corroboration of Eq. \eqref{eqn:drift_difussion_pd}}
Substituting Eq. \eqref{eqn:drift_difussion_pd} into the right hand side of Eq. \eqref{eqn:fourier_laplace_rearranged}, we find

\bea
\text{R.H.S.}&=&\frac{1+r\sum_{m=0}^\infty\frac{1}{r}\prod_{j=0}^m\frac{r}{ r-iVaka^j+Da^2k^2a^{2j}+s}}{r-iVk+Dk^2+s} \nn \\
&=&\frac{1+r\sum_{m=1}^\infty\frac{1}{r}\prod_{j=1}^m\frac{r}{ r-iVka^j+Dk^2a^{2j}+s}}{r-iVk+Dk^2+s} \nn \\
&=& \frac{1}{r-iVk+Dk^2+s} 
+ \sum_{m=1}^\infty\frac{1}{r}\prod_{j=0}^m\frac{r}{ r-iVka^j+Dk^2a^{2j}+s}\nn \\
&=&\sum_{m=0}^\infty\frac{1}{r}\prod_{j=0}^m\frac{r}{ r-iVka^j+Dk^2a^{2j}+s}~,
\eea
which is exactly $\Tilde{P}(k,s)$ in Eq. \eqref{eqn:drift_difussion_pd}.


\begin{thebibliography}{68}%
\makeatletter
\providecommand \@ifxundefined [1]{%
 \@ifx{#1\undefined}
}%
\providecommand \@ifnum [1]{%
 \ifnum #1\expandafter \@firstoftwo
 \else \expandafter \@secondoftwo
 \fi
}%
\providecommand \@ifx [1]{%
 \ifx #1\expandafter \@firstoftwo
 \else \expandafter \@secondoftwo
 \fi
}%
\providecommand \natexlab [1]{#1}%
\providecommand \enquote  [1]{``#1''}%
\providecommand \bibnamefont  [1]{#1}%
\providecommand \bibfnamefont [1]{#1}%
\providecommand \citenamefont [1]{#1}%
\providecommand \href@noop [0]{\@secondoftwo}%
\providecommand \href [0]{\begingroup \@sanitize@url \@href}%
\providecommand \@href[1]{\@@startlink{#1}\@@href}%
\providecommand \@@href[1]{\endgroup#1\@@endlink}%
\providecommand \@sanitize@url [0]{\catcode `\\12\catcode `\$12\catcode
  `\&12\catcode `\#12\catcode `\^12\catcode `\_12\catcode `\%12\relax}%
\providecommand \@@startlink[1]{}%
\providecommand \@@endlink[0]{}%
\providecommand \url  [0]{\begingroup\@sanitize@url \@url }%
\providecommand \@url [1]{\endgroup\@href {#1}{\urlprefix }}%
\providecommand \urlprefix  [0]{URL }%
\providecommand \Eprint [0]{\href }%
\providecommand \doibase [0]{https://doi.org/}%
\providecommand \selectlanguage [0]{\@gobble}%
\providecommand \bibinfo  [0]{\@secondoftwo}%
\providecommand \bibfield  [0]{\@secondoftwo}%
\providecommand \translation [1]{[#1]}%
\providecommand \BibitemOpen [0]{}%
\providecommand \bibitemStop [0]{}%
\providecommand \bibitemNoStop [0]{.\EOS\space}%
\providecommand \EOS [0]{\spacefactor3000\relax}%
\providecommand \BibitemShut  [1]{\csname bibitem#1\endcsname}%
\let\auto@bib@innerbib\@empty
\bibitem [{\citenamefont {Eliazar}\ \emph {et~al.}(2007)\citenamefont
  {Eliazar}, \citenamefont {Koren},\ and\ \citenamefont
  {Klafter}}]{eliazar_searching_2007}%
  \BibitemOpen
  \bibfield  {author} {\bibinfo {author} {\bibfnamefont {I.}~\bibnamefont
  {Eliazar}}, \bibinfo {author} {\bibfnamefont {T.}~\bibnamefont {Koren}},\
  and\ \bibinfo {author} {\bibfnamefont {J.}~\bibnamefont {Klafter}},\ }\href
  {https://doi.org/10.1088/0953-8984/19/6/065140} {\bibfield  {journal}
  {\bibinfo  {journal} {Journal of Physics: Condensed Matter}\ }\textbf
  {\bibinfo {volume} {19}},\ \bibinfo {pages} {065140} (\bibinfo {year}
  {2007})},\ \bibinfo {note} {publisher: IOP Publishing}\BibitemShut {NoStop}%
\bibitem [{\citenamefont {Kusmierz}\ \emph {et~al.}(2014)\citenamefont
  {Kusmierz}, \citenamefont {Majumdar}, \citenamefont {Sabhapandit},\ and\
  \citenamefont {Schehr}}]{kusmierz_first_2014}%
  \BibitemOpen
  \bibfield  {author} {\bibinfo {author} {\bibfnamefont {L.}~\bibnamefont
  {Kusmierz}}, \bibinfo {author} {\bibfnamefont {S.~N.}\ \bibnamefont
  {Majumdar}}, \bibinfo {author} {\bibfnamefont {S.}~\bibnamefont
  {Sabhapandit}},\ and\ \bibinfo {author} {\bibfnamefont {G.}~\bibnamefont
  {Schehr}},\ }\href {https://doi.org/10.1103/PhysRevLett.113.220602}
  {\bibfield  {journal} {\bibinfo  {journal} {Physical Review Letters}\
  }\textbf {\bibinfo {volume} {113}},\ \bibinfo {pages} {220602} (\bibinfo
  {year} {2014})},\ \bibinfo {note} {publisher: American Physical
  Society}\BibitemShut {NoStop}%
\bibitem [{\citenamefont {Ku{\'s}mierz}\ and\ \citenamefont
  {Gudowska-Nowak}(2015)}]{kusmierz_optimal_2015}%
  \BibitemOpen
  \bibfield  {author} {\bibinfo {author} {\bibfnamefont {{\L}.}~\bibnamefont
  {Ku{\'s}mierz}}\ and\ \bibinfo {author} {\bibfnamefont {E.}~\bibnamefont
  {Gudowska-Nowak}},\ }\href@noop {} {\bibfield  {journal} {\bibinfo  {journal}
  {Physical Review E}\ }\textbf {\bibinfo {volume} {92}},\ \bibinfo {pages}
  {052127} (\bibinfo {year} {2015})}\BibitemShut {NoStop}%
\bibitem [{\citenamefont {Rold{\'a}n}\ \emph {et~al.}(2016)\citenamefont
  {Rold{\'a}n}, \citenamefont {Lisica}, \citenamefont {S{\'a}nchez-Taltavull},\
  and\ \citenamefont {Grill}}]{roldan_stochastic_2016}%
  \BibitemOpen
  \bibfield  {author} {\bibinfo {author} {\bibfnamefont {{\'E}.}~\bibnamefont
  {Rold{\'a}n}}, \bibinfo {author} {\bibfnamefont {A.}~\bibnamefont {Lisica}},
  \bibinfo {author} {\bibfnamefont {D.}~\bibnamefont {S{\'a}nchez-Taltavull}},\
  and\ \bibinfo {author} {\bibfnamefont {S.~W.}\ \bibnamefont {Grill}},\
  }\href@noop {} {\bibfield  {journal} {\bibinfo  {journal} {Physical Review
  E}\ }\textbf {\bibinfo {volume} {93}},\ \bibinfo {pages} {062411} (\bibinfo
  {year} {2016})}\BibitemShut {NoStop}%
\bibitem [{\citenamefont {Nagar}\ and\ \citenamefont
  {Gupta}(2016)}]{nagar_diffusion_2016}%
  \BibitemOpen
  \bibfield  {author} {\bibinfo {author} {\bibfnamefont {A.}~\bibnamefont
  {Nagar}}\ and\ \bibinfo {author} {\bibfnamefont {S.}~\bibnamefont {Gupta}},\
  }\href {https://doi.org/10.1103/PhysRevE.93.060102} {\bibfield  {journal}
  {\bibinfo  {journal} {Physical Review E}\ }\textbf {\bibinfo {volume} {93}},\
  \bibinfo {pages} {060102} (\bibinfo {year} {2016})},\ \bibinfo {note}
  {publisher: American Physical Society}\BibitemShut {NoStop}%
\bibitem [{\citenamefont {Reuveni}(2016)}]{reuveni_optimal_2016}%
  \BibitemOpen
  \bibfield  {author} {\bibinfo {author} {\bibfnamefont {S.}~\bibnamefont
  {Reuveni}},\ }\href {https://doi.org/10.1103/PhysRevLett.116.170601}
  {\bibfield  {journal} {\bibinfo  {journal} {Physical Review Letters}\
  }\textbf {\bibinfo {volume} {116}},\ \bibinfo {pages} {170601} (\bibinfo
  {year} {2016})},\ \bibinfo {note} {publisher: American Physical
  Society}\BibitemShut {NoStop}%
\bibitem [{\citenamefont {Pal}\ and\ \citenamefont
  {Reuveni}(2017)}]{pal_first_2017}%
  \BibitemOpen
  \bibfield  {author} {\bibinfo {author} {\bibfnamefont {A.}~\bibnamefont
  {Pal}}\ and\ \bibinfo {author} {\bibfnamefont {S.}~\bibnamefont {Reuveni}},\
  }\href {https://doi.org/10.1103/PhysRevLett.118.030603} {\bibfield  {journal}
  {\bibinfo  {journal} {Physical Review Letters}\ }\textbf {\bibinfo {volume}
  {118}},\ \bibinfo {pages} {030603} (\bibinfo {year} {2017})},\ \bibinfo
  {note} {publisher: American Physical Society}\BibitemShut {NoStop}%
\bibitem [{\citenamefont {Falcao}\ and\ \citenamefont
  {Evans}(2017)}]{falcao_interacting_2017}%
  \BibitemOpen
  \bibfield  {author} {\bibinfo {author} {\bibfnamefont {R.}~\bibnamefont
  {Falcao}}\ and\ \bibinfo {author} {\bibfnamefont {M.~R.}\ \bibnamefont
  {Evans}},\ }\href {https://doi.org/10.1088/1742-5468/aa569c} {\bibfield
  {journal} {\bibinfo  {journal} {Journal of Statistical Mechanics: Theory and
  Experiment}\ }\textbf {\bibinfo {volume} {2017}},\ \bibinfo {pages} {023204}
  (\bibinfo {year} {2017})},\ \bibinfo {note} {publisher: IOP
  Publishing}\BibitemShut {NoStop}%
\bibitem [{\citenamefont {Montero}\ \emph {et~al.}(2017)\citenamefont
  {Montero}, \citenamefont {Masó-Puigdellosas},\ and\ \citenamefont
  {Villarroel}}]{montero_continuous-time_2017}%
  \BibitemOpen
  \bibfield  {author} {\bibinfo {author} {\bibfnamefont {M.}~\bibnamefont
  {Montero}}, \bibinfo {author} {\bibfnamefont {A.}~\bibnamefont
  {Masó-Puigdellosas}},\ and\ \bibinfo {author} {\bibfnamefont
  {J.}~\bibnamefont {Villarroel}},\ }\href
  {https://doi.org/10.1140/epjb/e2017-80348-4} {\bibfield  {journal} {\bibinfo
  {journal} {The European Physical Journal B}\ }\textbf {\bibinfo {volume}
  {90}},\ \bibinfo {pages} {176} (\bibinfo {year} {2017})}\BibitemShut
  {NoStop}%
\bibitem [{\citenamefont {Shkilev}(2017)}]{shkilev_continuous-time_2017}%
  \BibitemOpen
  \bibfield  {author} {\bibinfo {author} {\bibfnamefont {V.~P.}\ \bibnamefont
  {Shkilev}},\ }\href {https://doi.org/10.1103/PhysRevE.96.012126} {\bibfield
  {journal} {\bibinfo  {journal} {Physical Review E}\ }\textbf {\bibinfo
  {volume} {96}},\ \bibinfo {pages} {012126} (\bibinfo {year} {2017})},\
  \bibinfo {note} {publisher: American Physical Society}\BibitemShut {NoStop}%
\bibitem [{\citenamefont {Maes}\ and\ \citenamefont
  {Thiery}(2017)}]{maes_induced_2017}%
  \BibitemOpen
  \bibfield  {author} {\bibinfo {author} {\bibfnamefont {C.}~\bibnamefont
  {Maes}}\ and\ \bibinfo {author} {\bibfnamefont {T.}~\bibnamefont {Thiery}},\
  }\href {https://doi.org/10.1088/1751-8121/aa85a7} {\bibfield  {journal}
  {\bibinfo  {journal} {Journal of Physics A: Mathematical and Theoretical}\
  }\textbf {\bibinfo {volume} {50}},\ \bibinfo {pages} {415001} (\bibinfo
  {year} {2017})},\ \bibinfo {note} {publisher: IOP Publishing}\BibitemShut
  {NoStop}%
\bibitem [{\citenamefont {Evans}\ and\ \citenamefont
  {Majumdar}(2018)}]{evans_run_2018}%
  \BibitemOpen
  \bibfield  {author} {\bibinfo {author} {\bibfnamefont {M.~R.}\ \bibnamefont
  {Evans}}\ and\ \bibinfo {author} {\bibfnamefont {S.~N.}\ \bibnamefont
  {Majumdar}},\ }\href {https://doi.org/10.1088/1751-8121/aae74e} {\bibfield
  {journal} {\bibinfo  {journal} {Journal of Physics A: Mathematical and
  Theoretical}\ }\textbf {\bibinfo {volume} {51}},\ \bibinfo {pages} {475003}
  (\bibinfo {year} {2018})},\ \bibinfo {note} {publisher: IOP
  Publishing}\BibitemShut {NoStop}%
\bibitem [{\citenamefont {Chechkin}\ and\ \citenamefont
  {Sokolov}(2018)}]{chechkin_random_2018}%
  \BibitemOpen
  \bibfield  {author} {\bibinfo {author} {\bibfnamefont {A.}~\bibnamefont
  {Chechkin}}\ and\ \bibinfo {author} {\bibfnamefont {I.}~\bibnamefont
  {Sokolov}},\ }\href {https://doi.org/10.1103/PhysRevLett.121.050601}
  {\bibfield  {journal} {\bibinfo  {journal} {Physical Review Letters}\
  }\textbf {\bibinfo {volume} {121}},\ \bibinfo {pages} {050601} (\bibinfo
  {year} {2018})},\ \bibinfo {note} {publisher: American Physical
  Society}\BibitemShut {NoStop}%
\bibitem [{\citenamefont {Pal}\ \emph {et~al.}(2019{\natexlab{a}})\citenamefont
  {Pal}, \citenamefont {Eliazar},\ and\ \citenamefont
  {Reuveni}}]{pal_first_2019}%
  \BibitemOpen
  \bibfield  {author} {\bibinfo {author} {\bibfnamefont {A.}~\bibnamefont
  {Pal}}, \bibinfo {author} {\bibfnamefont {I.}~\bibnamefont {Eliazar}},\ and\
  \bibinfo {author} {\bibfnamefont {S.}~\bibnamefont {Reuveni}},\ }\href
  {https://doi.org/10.1103/PhysRevLett.122.020602} {\bibfield  {journal}
  {\bibinfo  {journal} {Physical Review Letters}\ }\textbf {\bibinfo {volume}
  {122}},\ \bibinfo {pages} {020602} (\bibinfo {year} {2019}{\natexlab{a}})},\
  \bibinfo {note} {publisher: American Physical Society}\BibitemShut {NoStop}%
\bibitem [{\citenamefont {Pal}\ \emph {et~al.}(2019{\natexlab{b}})\citenamefont
  {Pal}, \citenamefont {Chatterjee}, \citenamefont {Reuveni},\ and\
  \citenamefont {Kundu}}]{pal_local_2019}%
  \BibitemOpen
  \bibfield  {author} {\bibinfo {author} {\bibfnamefont {A.}~\bibnamefont
  {Pal}}, \bibinfo {author} {\bibfnamefont {R.}~\bibnamefont {Chatterjee}},
  \bibinfo {author} {\bibfnamefont {S.}~\bibnamefont {Reuveni}},\ and\ \bibinfo
  {author} {\bibfnamefont {A.}~\bibnamefont {Kundu}},\ }\href
  {https://doi.org/10.1088/1751-8121/ab2069} {\bibfield  {journal} {\bibinfo
  {journal} {Journal of Physics A: Mathematical and Theoretical}\ }\textbf
  {\bibinfo {volume} {52}},\ \bibinfo {pages} {264002} (\bibinfo {year}
  {2019}{\natexlab{b}})},\ \bibinfo {note} {publisher: IOP
  Publishing}\BibitemShut {NoStop}%
\bibitem [{\citenamefont {Santos}\ and\ \citenamefont
  {F}(2019)}]{santos_fractional_2019}%
  \BibitemOpen
  \bibfield  {author} {\bibinfo {author} {\bibfnamefont {D.}~\bibnamefont
  {Santos}}\ and\ \bibinfo {author} {\bibfnamefont {M.~A.}\ \bibnamefont {F}},\
  }\href {https://doi.org/10.3390/physics1010005} {\bibfield  {journal}
  {\bibinfo  {journal} {Physics}\ }\textbf {\bibinfo {volume} {1}},\ \bibinfo
  {pages} {40} (\bibinfo {year} {2019})},\ \bibinfo {note} {number: 1
  Publisher: Multidisciplinary Digital Publishing Institute}\BibitemShut
  {NoStop}%
\bibitem [{\citenamefont {Ahmad}\ \emph {et~al.}(2019)\citenamefont {Ahmad},
  \citenamefont {Nayak}, \citenamefont {Bansal}, \citenamefont {Nandi},\ and\
  \citenamefont {Das}}]{ahmad_first_2019}%
  \BibitemOpen
  \bibfield  {author} {\bibinfo {author} {\bibfnamefont {S.}~\bibnamefont
  {Ahmad}}, \bibinfo {author} {\bibfnamefont {I.}~\bibnamefont {Nayak}},
  \bibinfo {author} {\bibfnamefont {A.}~\bibnamefont {Bansal}}, \bibinfo
  {author} {\bibfnamefont {A.}~\bibnamefont {Nandi}},\ and\ \bibinfo {author}
  {\bibfnamefont {D.}~\bibnamefont {Das}},\ }\href
  {https://doi.org/10.1103/PhysRevE.99.022130} {\bibfield  {journal} {\bibinfo
  {journal} {Physical Review E}\ }\textbf {\bibinfo {volume} {99}},\ \bibinfo
  {pages} {022130} (\bibinfo {year} {2019})},\ \bibinfo {note} {publisher:
  American Physical Society}\BibitemShut {NoStop}%
\bibitem [{\citenamefont {Ray}\ \emph {et~al.}(2019)\citenamefont {Ray},
  \citenamefont {Mondal},\ and\ \citenamefont {Reuveni}}]{ray_peclet_2019}%
  \BibitemOpen
  \bibfield  {author} {\bibinfo {author} {\bibfnamefont {S.}~\bibnamefont
  {Ray}}, \bibinfo {author} {\bibfnamefont {D.}~\bibnamefont {Mondal}},\ and\
  \bibinfo {author} {\bibfnamefont {S.}~\bibnamefont {Reuveni}},\ }\href
  {https://doi.org/10.1088/1751-8121/ab1fcc} {\bibfield  {journal} {\bibinfo
  {journal} {Journal of Physics A: Mathematical and Theoretical}\ }\textbf
  {\bibinfo {volume} {52}},\ \bibinfo {pages} {255002} (\bibinfo {year}
  {2019})},\ \bibinfo {note} {publisher: IOP Publishing}\BibitemShut {NoStop}%
\bibitem [{\citenamefont {Bodrova}\ \emph
  {et~al.}(2019{\natexlab{a}})\citenamefont {Bodrova}, \citenamefont
  {Chechkin},\ and\ \citenamefont {Sokolov}}]{bodrova_nonrenewal_2019}%
  \BibitemOpen
  \bibfield  {author} {\bibinfo {author} {\bibfnamefont {A.~S.}\ \bibnamefont
  {Bodrova}}, \bibinfo {author} {\bibfnamefont {A.~V.}\ \bibnamefont
  {Chechkin}},\ and\ \bibinfo {author} {\bibfnamefont {I.~M.}\ \bibnamefont
  {Sokolov}},\ }\href {https://doi.org/10.1103/PhysRevE.100.012119} {\bibfield
  {journal} {\bibinfo  {journal} {Physical Review E}\ }\textbf {\bibinfo
  {volume} {100}},\ \bibinfo {pages} {012119} (\bibinfo {year}
  {2019}{\natexlab{a}})},\ \bibinfo {note} {publisher: American Physical
  Society}\BibitemShut {NoStop}%
\bibitem [{\citenamefont {Masó-Puigdellosas}\ \emph
  {et~al.}(2019)\citenamefont {Masó-Puigdellosas}, \citenamefont {Campos},\
  and\ \citenamefont {Méndez}}]{maso-puigdellosas_transport_2019}%
  \BibitemOpen
  \bibfield  {author} {\bibinfo {author} {\bibfnamefont {A.}~\bibnamefont
  {Masó-Puigdellosas}}, \bibinfo {author} {\bibfnamefont {D.}~\bibnamefont
  {Campos}},\ and\ \bibinfo {author} {\bibfnamefont {V.}~\bibnamefont
  {Méndez}},\ }\href {https://doi.org/10.1103/PhysRevE.100.042104} {\bibfield
  {journal} {\bibinfo  {journal} {Physical Review E}\ }\textbf {\bibinfo
  {volume} {100}},\ \bibinfo {pages} {042104} (\bibinfo {year} {2019})},\
  \bibinfo {note} {publisher: American Physical Society}\BibitemShut {NoStop}%
\bibitem [{\citenamefont {Pal}\ \emph {et~al.}(2020)\citenamefont {Pal},
  \citenamefont {Ku{\'s}mierz},\ and\ \citenamefont
  {Reuveni}}]{pal_search_2020}%
  \BibitemOpen
  \bibfield  {author} {\bibinfo {author} {\bibfnamefont {A.}~\bibnamefont
  {Pal}}, \bibinfo {author} {\bibfnamefont {{\L}.}~\bibnamefont
  {Ku{\'s}mierz}},\ and\ \bibinfo {author} {\bibfnamefont {S.}~\bibnamefont
  {Reuveni}},\ }\href@noop {} {\bibfield  {journal} {\bibinfo  {journal}
  {Physical Review Research}\ }\textbf {\bibinfo {volume} {2}},\ \bibinfo
  {pages} {043174} (\bibinfo {year} {2020})}\BibitemShut {NoStop}%
\bibitem [{\citenamefont {Ray}\ and\ \citenamefont
  {Reuveni}(2020)}]{ray_diffusion_2020}%
  \BibitemOpen
  \bibfield  {author} {\bibinfo {author} {\bibfnamefont {S.}~\bibnamefont
  {Ray}}\ and\ \bibinfo {author} {\bibfnamefont {S.}~\bibnamefont {Reuveni}},\
  }\href {https://doi.org/10.1063/5.0010549} {\bibfield  {journal} {\bibinfo
  {journal} {The Journal of Chemical Physics}\ }\textbf {\bibinfo {volume}
  {152}},\ \bibinfo {pages} {234110} (\bibinfo {year} {2020})},\ \bibinfo
  {note} {publisher: American Institute of Physics}\BibitemShut {NoStop}%
\bibitem [{\citenamefont {Riascos}\ \emph {et~al.}(2020)\citenamefont
  {Riascos}, \citenamefont {Boyer}, \citenamefont {Herringer},\ and\
  \citenamefont {Mateos}}]{riascos_random_2020}%
  \BibitemOpen
  \bibfield  {author} {\bibinfo {author} {\bibfnamefont {A.~P.}\ \bibnamefont
  {Riascos}}, \bibinfo {author} {\bibfnamefont {D.}~\bibnamefont {Boyer}},
  \bibinfo {author} {\bibfnamefont {P.}~\bibnamefont {Herringer}},\ and\
  \bibinfo {author} {\bibfnamefont {J.~L.}\ \bibnamefont {Mateos}},\ }\href
  {https://doi.org/10.1103/PhysRevE.101.062147} {\bibfield  {journal} {\bibinfo
   {journal} {Physical Review E}\ }\textbf {\bibinfo {volume} {101}},\ \bibinfo
  {pages} {062147} (\bibinfo {year} {2020})},\ \bibinfo {note} {publisher:
  American Physical Society}\BibitemShut {NoStop}%
\bibitem [{\citenamefont
  {Bressloff}(2020{\natexlab{a}})}]{bressloff_modeling_2020}%
  \BibitemOpen
  \bibfield  {author} {\bibinfo {author} {\bibfnamefont {P.~C.}\ \bibnamefont
  {Bressloff}},\ }\href {https://doi.org/10.1088/1751-8121/ab9fb7} {\bibfield
  {journal} {\bibinfo  {journal} {Journal of Physics A: Mathematical and
  Theoretical}\ }\textbf {\bibinfo {volume} {53}},\ \bibinfo {pages} {355001}
  (\bibinfo {year} {2020}{\natexlab{a}})},\ \bibinfo {note} {publisher: IOP
  Publishing}\BibitemShut {NoStop}%
\bibitem [{\citenamefont {Pinsky}(2020)}]{pinsky_diffusive_2020}%
  \BibitemOpen
  \bibfield  {author} {\bibinfo {author} {\bibfnamefont {R.~G.}\ \bibnamefont
  {Pinsky}},\ }\href {https://doi.org/10.1016/j.spa.2019.08.008} {\bibfield
  {journal} {\bibinfo  {journal} {Stochastic Processes and their Applications}\
  }\textbf {\bibinfo {volume} {130}},\ \bibinfo {pages} {2954} (\bibinfo {year}
  {2020})}\BibitemShut {NoStop}%
\bibitem [{\citenamefont
  {Bressloff}(2020{\natexlab{b}})}]{bressloff_directed_2020}%
  \BibitemOpen
  \bibfield  {author} {\bibinfo {author} {\bibfnamefont {P.~C.}\ \bibnamefont
  {Bressloff}},\ }\href {https://doi.org/10.1088/1751-8121/ab7138} {\bibfield
  {journal} {\bibinfo  {journal} {Journal of Physics A: Mathematical and
  Theoretical}\ }\textbf {\bibinfo {volume} {53}},\ \bibinfo {pages} {105001}
  (\bibinfo {year} {2020}{\natexlab{b}})},\ \bibinfo {note} {publisher: IOP
  Publishing}\BibitemShut {NoStop}%
\bibitem [{\citenamefont {De~Bruyne}\ \emph {et~al.}(2020)\citenamefont
  {De~Bruyne}, \citenamefont {Randon-Furling},\ and\ \citenamefont
  {Redner}}]{de_bruyne_optimization_2020}%
  \BibitemOpen
  \bibfield  {author} {\bibinfo {author} {\bibfnamefont {B.}~\bibnamefont
  {De~Bruyne}}, \bibinfo {author} {\bibfnamefont {J.}~\bibnamefont
  {Randon-Furling}},\ and\ \bibinfo {author} {\bibfnamefont {S.}~\bibnamefont
  {Redner}},\ }\href {https://doi.org/10.1103/PhysRevLett.125.050602}
  {\bibfield  {journal} {\bibinfo  {journal} {Physical Review Letters}\
  }\textbf {\bibinfo {volume} {125}},\ \bibinfo {pages} {050602} (\bibinfo
  {year} {2020})},\ \bibinfo {note} {publisher: American Physical
  Society}\BibitemShut {NoStop}%
\bibitem [{\citenamefont {Bodrova}\ and\ \citenamefont
  {Sokolov}(2020{\natexlab{a}})}]{bodrova_continuous-time_2020}%
  \BibitemOpen
  \bibfield  {author} {\bibinfo {author} {\bibfnamefont {A.~S.}\ \bibnamefont
  {Bodrova}}\ and\ \bibinfo {author} {\bibfnamefont {I.~M.}\ \bibnamefont
  {Sokolov}},\ }\href {https://doi.org/10.1103/PhysRevE.101.062117} {\bibfield
  {journal} {\bibinfo  {journal} {Physical Review E}\ }\textbf {\bibinfo
  {volume} {101}},\ \bibinfo {pages} {062117} (\bibinfo {year}
  {2020}{\natexlab{a}})},\ \bibinfo {note} {publisher: American Physical
  Society}\BibitemShut {NoStop}%
\bibitem [{\citenamefont {González}\ \emph {et~al.}(2021)\citenamefont
  {González}, \citenamefont {Riascos},\ and\ \citenamefont
  {Boyer}}]{gonzalez_diffusive_2021}%
  \BibitemOpen
  \bibfield  {author} {\bibinfo {author} {\bibfnamefont {F.~H.}\ \bibnamefont
  {González}}, \bibinfo {author} {\bibfnamefont {A.~P.}\ \bibnamefont
  {Riascos}},\ and\ \bibinfo {author} {\bibfnamefont {D.}~\bibnamefont
  {Boyer}},\ }\href {https://doi.org/10.1103/PhysRevE.103.062126} {\bibfield
  {journal} {\bibinfo  {journal} {Physical Review E}\ }\textbf {\bibinfo
  {volume} {103}},\ \bibinfo {pages} {062126} (\bibinfo {year} {2021})},\
  \bibinfo {note} {publisher: American Physical Society}\BibitemShut {NoStop}%
\bibitem [{\citenamefont {Singh}\ \emph {et~al.}(2020)\citenamefont {Singh},
  \citenamefont {Metzler},\ and\ \citenamefont
  {Sandev}}]{singh_resetting_2020}%
  \BibitemOpen
  \bibfield  {author} {\bibinfo {author} {\bibfnamefont {R.~K.}\ \bibnamefont
  {Singh}}, \bibinfo {author} {\bibfnamefont {R.}~\bibnamefont {Metzler}},\
  and\ \bibinfo {author} {\bibfnamefont {T.}~\bibnamefont {Sandev}},\ }\href
  {https://doi.org/10.1088/1751-8121/abc83a} {\bibfield  {journal} {\bibinfo
  {journal} {Journal of Physics A: Mathematical and Theoretical}\ }\textbf
  {\bibinfo {volume} {53}},\ \bibinfo {pages} {505003} (\bibinfo {year}
  {2020})},\ \bibinfo {note} {publisher: IOP Publishing}\BibitemShut {NoStop}%
\bibitem [{\citenamefont {Stojkoski}\ \emph {et~al.}(2022)\citenamefont
  {Stojkoski}, \citenamefont {Jolakoski}, \citenamefont {Pal}, \citenamefont
  {Sandev}, \citenamefont {Kocarev},\ and\ \citenamefont
  {Metzler}}]{stojkoski2022income}%
  \BibitemOpen
  \bibfield  {author} {\bibinfo {author} {\bibfnamefont {V.}~\bibnamefont
  {Stojkoski}}, \bibinfo {author} {\bibfnamefont {P.}~\bibnamefont
  {Jolakoski}}, \bibinfo {author} {\bibfnamefont {A.}~\bibnamefont {Pal}},
  \bibinfo {author} {\bibfnamefont {T.}~\bibnamefont {Sandev}}, \bibinfo
  {author} {\bibfnamefont {L.}~\bibnamefont {Kocarev}},\ and\ \bibinfo {author}
  {\bibfnamefont {R.}~\bibnamefont {Metzler}},\ }\href@noop {} {\bibfield
  {journal} {\bibinfo  {journal} {Philosophical Transactions of the Royal
  Society A}\ }\textbf {\bibinfo {volume} {380}},\ \bibinfo {pages} {20210157}
  (\bibinfo {year} {2022})}\BibitemShut {NoStop}%
\bibitem [{\citenamefont {Vinod}\ \emph {et~al.}(2022)\citenamefont {Vinod},
  \citenamefont {Cherstvy}, \citenamefont {Wang}, \citenamefont {Metzler},\
  and\ \citenamefont {Sokolov}}]{vinod2022nonergodicity}%
  \BibitemOpen
  \bibfield  {author} {\bibinfo {author} {\bibfnamefont {D.}~\bibnamefont
  {Vinod}}, \bibinfo {author} {\bibfnamefont {A.~G.}\ \bibnamefont {Cherstvy}},
  \bibinfo {author} {\bibfnamefont {W.}~\bibnamefont {Wang}}, \bibinfo {author}
  {\bibfnamefont {R.}~\bibnamefont {Metzler}},\ and\ \bibinfo {author}
  {\bibfnamefont {I.~M.}\ \bibnamefont {Sokolov}},\ }\href@noop {} {\bibfield
  {journal} {\bibinfo  {journal} {Physical Review E}\ }\textbf {\bibinfo
  {volume} {105}},\ \bibinfo {pages} {L012106} (\bibinfo {year}
  {2022})}\BibitemShut {NoStop}%
\bibitem [{\citenamefont {Wang}\ \emph {et~al.}(2021)\citenamefont {Wang},
  \citenamefont {Cherstvy}, \citenamefont {Kantz}, \citenamefont {Metzler},\
  and\ \citenamefont {Sokolov}}]{PhysRevE.104.024105}%
  \BibitemOpen
  \bibfield  {author} {\bibinfo {author} {\bibfnamefont {W.}~\bibnamefont
  {Wang}}, \bibinfo {author} {\bibfnamefont {A.~G.}\ \bibnamefont {Cherstvy}},
  \bibinfo {author} {\bibfnamefont {H.}~\bibnamefont {Kantz}}, \bibinfo
  {author} {\bibfnamefont {R.}~\bibnamefont {Metzler}},\ and\ \bibinfo {author}
  {\bibfnamefont {I.~M.}\ \bibnamefont {Sokolov}},\ }\href
  {https://doi.org/10.1103/PhysRevE.104.024105} {\bibfield  {journal} {\bibinfo
   {journal} {Phys. Rev. E}\ }\textbf {\bibinfo {volume} {104}},\ \bibinfo
  {pages} {024105} (\bibinfo {year} {2021})}\BibitemShut {NoStop}%
\bibitem [{\citenamefont {Tal-Friedman}\ \emph {et~al.}(2020)\citenamefont
  {Tal-Friedman}, \citenamefont {Pal}, \citenamefont {Sekhon}, \citenamefont
  {Reuveni},\ and\ \citenamefont {Roichman}}]{tal-friedman_experimental_2020}%
  \BibitemOpen
  \bibfield  {author} {\bibinfo {author} {\bibfnamefont {O.}~\bibnamefont
  {Tal-Friedman}}, \bibinfo {author} {\bibfnamefont {A.}~\bibnamefont {Pal}},
  \bibinfo {author} {\bibfnamefont {A.}~\bibnamefont {Sekhon}}, \bibinfo
  {author} {\bibfnamefont {S.}~\bibnamefont {Reuveni}},\ and\ \bibinfo {author}
  {\bibfnamefont {Y.}~\bibnamefont {Roichman}},\ }\href
  {https://doi.org/10.1021/acs.jpclett.0c02122} {\bibfield  {journal} {\bibinfo
   {journal} {The Journal of Physical Chemistry Letters}\ }\textbf {\bibinfo
  {volume} {11}},\ \bibinfo {pages} {7350} (\bibinfo {year} {2020})},\ \bibinfo
  {note} {publisher: American Chemical Society}\BibitemShut {NoStop}%
\bibitem [{\citenamefont {Besga}\ \emph {et~al.}(2020)\citenamefont {Besga},
  \citenamefont {Bovon}, \citenamefont {Petrosyan}, \citenamefont {Majumdar},\
  and\ \citenamefont {Ciliberto}}]{besga_optimal_2020}%
  \BibitemOpen
  \bibfield  {author} {\bibinfo {author} {\bibfnamefont {B.}~\bibnamefont
  {Besga}}, \bibinfo {author} {\bibfnamefont {A.}~\bibnamefont {Bovon}},
  \bibinfo {author} {\bibfnamefont {A.}~\bibnamefont {Petrosyan}}, \bibinfo
  {author} {\bibfnamefont {S.~N.}\ \bibnamefont {Majumdar}},\ and\ \bibinfo
  {author} {\bibfnamefont {S.}~\bibnamefont {Ciliberto}},\ }\href
  {https://doi.org/10.1103/PhysRevResearch.2.032029} {\bibfield  {journal}
  {\bibinfo  {journal} {Physical Review Research}\ }\textbf {\bibinfo {volume}
  {2}},\ \bibinfo {pages} {032029} (\bibinfo {year} {2020})},\ \bibinfo {note}
  {publisher: American Physical Society}\BibitemShut {NoStop}%
\bibitem [{\citenamefont {Faisant}\ \emph {et~al.}(2021)\citenamefont
  {Faisant}, \citenamefont {Besga}, \citenamefont {Petrosyan}, \citenamefont
  {Ciliberto},\ and\ \citenamefont {Majumdar}}]{faisant_optimal_2021}%
  \BibitemOpen
  \bibfield  {author} {\bibinfo {author} {\bibfnamefont {F.}~\bibnamefont
  {Faisant}}, \bibinfo {author} {\bibfnamefont {B.}~\bibnamefont {Besga}},
  \bibinfo {author} {\bibfnamefont {A.}~\bibnamefont {Petrosyan}}, \bibinfo
  {author} {\bibfnamefont {S.}~\bibnamefont {Ciliberto}},\ and\ \bibinfo
  {author} {\bibfnamefont {S.~N.}\ \bibnamefont {Majumdar}},\ }\href
  {https://doi.org/10.1088/1742-5468/ac2cc7} {\bibfield  {journal} {\bibinfo
  {journal} {Journal of Statistical Mechanics: Theory and Experiment}\ }\textbf
  {\bibinfo {volume} {2021}},\ \bibinfo {pages} {113203} (\bibinfo {year}
  {2021})},\ \bibinfo {note} {publisher: IOP Publishing}\BibitemShut {NoStop}%
\bibitem [{\citenamefont {Evans}\ and\ \citenamefont
  {Majumdar}(2011)}]{evans_diffusion_2011}%
  \BibitemOpen
  \bibfield  {author} {\bibinfo {author} {\bibfnamefont {M.~R.}\ \bibnamefont
  {Evans}}\ and\ \bibinfo {author} {\bibfnamefont {S.~N.}\ \bibnamefont
  {Majumdar}},\ }\href {https://doi.org/10.1103/PhysRevLett.106.160601}
  {\bibfield  {journal} {\bibinfo  {journal} {Physical Review Letters}\
  }\textbf {\bibinfo {volume} {106}},\ \bibinfo {pages} {160601} (\bibinfo
  {year} {2011})},\ \bibinfo {note} {publisher: American Physical
  Society}\BibitemShut {NoStop}%
\bibitem [{\citenamefont {Evans}\ and\ \citenamefont
  {Majumdar}(2014)}]{evans_diffusion_2014}%
  \BibitemOpen
  \bibfield  {author} {\bibinfo {author} {\bibfnamefont {M.~R.}\ \bibnamefont
  {Evans}}\ and\ \bibinfo {author} {\bibfnamefont {S.~N.}\ \bibnamefont
  {Majumdar}},\ }\href {https://doi.org/10.1088/1751-8113/47/28/285001}
  {\bibfield  {journal} {\bibinfo  {journal} {Journal of Physics A:
  Mathematical and Theoretical}\ }\textbf {\bibinfo {volume} {47}},\ \bibinfo
  {pages} {285001} (\bibinfo {year} {2014})},\ \bibinfo {note} {publisher: IOP
  Publishing}\BibitemShut {NoStop}%
\bibitem [{\citenamefont {Pal}(2015)}]{pal_diffusion_2015}%
  \BibitemOpen
  \bibfield  {author} {\bibinfo {author} {\bibfnamefont {A.}~\bibnamefont
  {Pal}},\ }\href {https://doi.org/10.1103/PhysRevE.91.012113} {\bibfield
  {journal} {\bibinfo  {journal} {Physical Review E}\ }\textbf {\bibinfo
  {volume} {91}},\ \bibinfo {pages} {012113} (\bibinfo {year} {2015})},\
  \bibinfo {note} {publisher: American Physical Society}\BibitemShut {NoStop}%
\bibitem [{\citenamefont {Pal}\ \emph {et~al.}(2016)\citenamefont {Pal},
  \citenamefont {Kundu},\ and\ \citenamefont {Evans}}]{pal_diffusion_2016}%
  \BibitemOpen
  \bibfield  {author} {\bibinfo {author} {\bibfnamefont {A.}~\bibnamefont
  {Pal}}, \bibinfo {author} {\bibfnamefont {A.}~\bibnamefont {Kundu}},\ and\
  \bibinfo {author} {\bibfnamefont {M.~R.}\ \bibnamefont {Evans}},\ }\href
  {https://doi.org/10.1088/1751-8113/49/22/225001} {\bibfield  {journal}
  {\bibinfo  {journal} {Journal of Physics A: Mathematical and Theoretical}\
  }\textbf {\bibinfo {volume} {49}},\ \bibinfo {pages} {225001} (\bibinfo
  {year} {2016})},\ \bibinfo {note} {publisher: IOP Publishing}\BibitemShut
  {NoStop}%
\bibitem [{\citenamefont {Méndez}\ and\ \citenamefont
  {Campos}(2016)}]{mendez_characterization_2016}%
  \BibitemOpen
  \bibfield  {author} {\bibinfo {author} {\bibfnamefont {V.}~\bibnamefont
  {Méndez}}\ and\ \bibinfo {author} {\bibfnamefont {D.}~\bibnamefont
  {Campos}},\ }\href {https://doi.org/10.1103/PhysRevE.93.022106} {\bibfield
  {journal} {\bibinfo  {journal} {Physical Review E}\ }\textbf {\bibinfo
  {volume} {93}},\ \bibinfo {pages} {022106} (\bibinfo {year} {2016})},\
  \bibinfo {note} {publisher: American Physical Society}\BibitemShut {NoStop}%
\bibitem [{\citenamefont {Montero}\ and\ \citenamefont
  {Villarroel}(2016)}]{montero_directed_2016}%
  \BibitemOpen
  \bibfield  {author} {\bibinfo {author} {\bibfnamefont {M.}~\bibnamefont
  {Montero}}\ and\ \bibinfo {author} {\bibfnamefont {J.}~\bibnamefont
  {Villarroel}},\ }\href {https://doi.org/10.1103/PhysRevE.94.032132}
  {\bibfield  {journal} {\bibinfo  {journal} {Physical Review E}\ }\textbf
  {\bibinfo {volume} {94}},\ \bibinfo {pages} {032132} (\bibinfo {year}
  {2016})},\ \bibinfo {note} {publisher: American Physical Society}\BibitemShut
  {NoStop}%
\bibitem [{\citenamefont {Eule}\ and\ \citenamefont
  {Metzger}(2016)}]{eule_non-equilibrium_2016}%
  \BibitemOpen
  \bibfield  {author} {\bibinfo {author} {\bibfnamefont {S.}~\bibnamefont
  {Eule}}\ and\ \bibinfo {author} {\bibfnamefont {J.~J.}\ \bibnamefont
  {Metzger}},\ }\href {https://doi.org/10.1088/1367-2630/18/3/033006}
  {\bibfield  {journal} {\bibinfo  {journal} {New Journal of Physics}\ }\textbf
  {\bibinfo {volume} {18}},\ \bibinfo {pages} {033006} (\bibinfo {year}
  {2016})},\ \bibinfo {note} {publisher: IOP Publishing}\BibitemShut {NoStop}%
\bibitem [{\citenamefont {Pal}\ \emph {et~al.}(2019{\natexlab{c}})\citenamefont
  {Pal}, \citenamefont {Ku{\'s}mierz},\ and\ \citenamefont
  {Reuveni}}]{pal_invariants_2019}%
  \BibitemOpen
  \bibfield  {author} {\bibinfo {author} {\bibfnamefont {A.}~\bibnamefont
  {Pal}}, \bibinfo {author} {\bibfnamefont {{\L}.}~\bibnamefont
  {Ku{\'s}mierz}},\ and\ \bibinfo {author} {\bibfnamefont {S.}~\bibnamefont
  {Reuveni}},\ }\href@noop {} {\bibfield  {journal} {\bibinfo  {journal} {New
  Journal of Physics}\ }\textbf {\bibinfo {volume} {21}},\ \bibinfo {pages}
  {113024} (\bibinfo {year} {2019}{\natexlab{c}})}\BibitemShut {NoStop}%
\bibitem [{\citenamefont {Masoliver}(2019)}]{masoliver_telegraphic_2019}%
  \BibitemOpen
  \bibfield  {author} {\bibinfo {author} {\bibfnamefont {J.}~\bibnamefont
  {Masoliver}},\ }\href {https://doi.org/10.1103/PhysRevE.99.012121} {\bibfield
   {journal} {\bibinfo  {journal} {Physical Review E}\ }\textbf {\bibinfo
  {volume} {99}},\ \bibinfo {pages} {012121} (\bibinfo {year} {2019})},\
  \bibinfo {note} {publisher: American Physical Society}\BibitemShut {NoStop}%
\bibitem [{\citenamefont {Bodrova}\ \emph
  {et~al.}(2019{\natexlab{b}})\citenamefont {Bodrova}, \citenamefont
  {Chechkin},\ and\ \citenamefont {Sokolov}}]{bodrova_scaled_2019}%
  \BibitemOpen
  \bibfield  {author} {\bibinfo {author} {\bibfnamefont {A.~S.}\ \bibnamefont
  {Bodrova}}, \bibinfo {author} {\bibfnamefont {A.~V.}\ \bibnamefont
  {Chechkin}},\ and\ \bibinfo {author} {\bibfnamefont {I.~M.}\ \bibnamefont
  {Sokolov}},\ }\href {https://doi.org/10.1103/PhysRevE.100.012120} {\bibfield
  {journal} {\bibinfo  {journal} {Physical Review E}\ }\textbf {\bibinfo
  {volume} {100}},\ \bibinfo {pages} {012120} (\bibinfo {year}
  {2019}{\natexlab{b}})},\ \bibinfo {note} {publisher: American Physical
  Society}\BibitemShut {NoStop}%
\bibitem [{\citenamefont {Pal}\ \emph {et~al.}(2019{\natexlab{d}})\citenamefont
  {Pal}, \citenamefont {Ku{\'s}mierz},\ and\ \citenamefont
  {Reuveni}}]{pal_time-dependent_2019}%
  \BibitemOpen
  \bibfield  {author} {\bibinfo {author} {\bibfnamefont {A.}~\bibnamefont
  {Pal}}, \bibinfo {author} {\bibfnamefont {{\L}.}~\bibnamefont
  {Ku{\'s}mierz}},\ and\ \bibinfo {author} {\bibfnamefont {S.}~\bibnamefont
  {Reuveni}},\ }\href@noop {} {\bibfield  {journal} {\bibinfo  {journal}
  {Physical Review E}\ }\textbf {\bibinfo {volume} {100}},\ \bibinfo {pages}
  {040101} (\bibinfo {year} {2019}{\natexlab{d}})}\BibitemShut {NoStop}%
\bibitem [{\citenamefont {Ku{\'s}mierz}\ and\ \citenamefont
  {Gudowska-Nowak}(2019)}]{kusmierz_subdiffusive_2019}%
  \BibitemOpen
  \bibfield  {author} {\bibinfo {author} {\bibfnamefont {{\L}.}~\bibnamefont
  {Ku{\'s}mierz}}\ and\ \bibinfo {author} {\bibfnamefont {E.}~\bibnamefont
  {Gudowska-Nowak}},\ }\href@noop {} {\bibfield  {journal} {\bibinfo  {journal}
  {Physical Review E}\ }\textbf {\bibinfo {volume} {99}},\ \bibinfo {pages}
  {052116} (\bibinfo {year} {2019})}\BibitemShut {NoStop}%
\bibitem [{\citenamefont {Gupta}(2019)}]{gupta_stochastic_2019}%
  \BibitemOpen
  \bibfield  {author} {\bibinfo {author} {\bibfnamefont {D.}~\bibnamefont
  {Gupta}},\ }\href {https://doi.org/10.1088/1742-5468/ab054a} {\bibfield
  {journal} {\bibinfo  {journal} {Journal of Statistical Mechanics: Theory and
  Experiment}\ }\textbf {\bibinfo {volume} {2019}},\ \bibinfo {pages} {033212}
  (\bibinfo {year} {2019})},\ \bibinfo {note} {publisher: IOP
  Publishing}\BibitemShut {NoStop}%
\bibitem [{\citenamefont {Bodrova}\ and\ \citenamefont
  {Sokolov}(2020{\natexlab{b}})}]{bodrova_resetting_2020}%
  \BibitemOpen
  \bibfield  {author} {\bibinfo {author} {\bibfnamefont {A.~S.}\ \bibnamefont
  {Bodrova}}\ and\ \bibinfo {author} {\bibfnamefont {I.~M.}\ \bibnamefont
  {Sokolov}},\ }\href {https://doi.org/10.1103/PhysRevE.101.052130} {\bibfield
  {journal} {\bibinfo  {journal} {Physical Review E}\ }\textbf {\bibinfo
  {volume} {101}},\ \bibinfo {pages} {052130} (\bibinfo {year}
  {2020}{\natexlab{b}})},\ \bibinfo {note} {publisher: American Physical
  Society}\BibitemShut {NoStop}%
\bibitem [{\citenamefont {Bodrova}\ and\ \citenamefont
  {Sokolov}(2020{\natexlab{c}})}]{bodrova_brownian_2020}%
  \BibitemOpen
  \bibfield  {author} {\bibinfo {author} {\bibfnamefont {A.~S.}\ \bibnamefont
  {Bodrova}}\ and\ \bibinfo {author} {\bibfnamefont {I.~M.}\ \bibnamefont
  {Sokolov}},\ }\href {https://doi.org/10.1103/PhysRevE.102.032129} {\bibfield
  {journal} {\bibinfo  {journal} {Physical Review E}\ }\textbf {\bibinfo
  {volume} {102}},\ \bibinfo {pages} {032129} (\bibinfo {year}
  {2020}{\natexlab{c}})},\ \bibinfo {note} {publisher: American Physical
  Society}\BibitemShut {NoStop}%
\bibitem [{\citenamefont {Evans}\ \emph {et~al.}(2020)\citenamefont {Evans},
  \citenamefont {Majumdar},\ and\ \citenamefont
  {Schehr}}]{evans_stochastic_2020}%
  \BibitemOpen
  \bibfield  {author} {\bibinfo {author} {\bibfnamefont {M.~R.}\ \bibnamefont
  {Evans}}, \bibinfo {author} {\bibfnamefont {S.~N.}\ \bibnamefont
  {Majumdar}},\ and\ \bibinfo {author} {\bibfnamefont {G.}~\bibnamefont
  {Schehr}},\ }\href {https://doi.org/10.1088/1751-8121/ab7cfe} {\bibfield
  {journal} {\bibinfo  {journal} {Journal of Physics A: Mathematical and
  Theoretical}\ }\textbf {\bibinfo {volume} {53}},\ \bibinfo {pages} {193001}
  (\bibinfo {year} {2020})},\ \bibinfo {note} {publisher: IOP
  Publishing}\BibitemShut {NoStop}%
\bibitem [{\citenamefont {Miron}\ and\ \citenamefont
  {Reuveni}(2021)}]{miron_diffusion_2021}%
  \BibitemOpen
  \bibfield  {author} {\bibinfo {author} {\bibfnamefont {A.}~\bibnamefont
  {Miron}}\ and\ \bibinfo {author} {\bibfnamefont {S.}~\bibnamefont
  {Reuveni}},\ }\href {https://doi.org/10.1103/PhysRevResearch.3.L012023}
  {\bibfield  {journal} {\bibinfo  {journal} {Physical Review Research}\
  }\textbf {\bibinfo {volume} {3}},\ \bibinfo {pages} {L012023} (\bibinfo
  {year} {2021})},\ \bibinfo {note} {publisher: American Physical
  Society}\BibitemShut {NoStop}%
\bibitem [{\citenamefont {Stojkoski}\ \emph {et~al.}(2021)\citenamefont
  {Stojkoski}, \citenamefont {Sandev}, \citenamefont {Kocarev},\ and\
  \citenamefont {Pal}}]{PhysRevE.104.014121}%
  \BibitemOpen
  \bibfield  {author} {\bibinfo {author} {\bibfnamefont {V.}~\bibnamefont
  {Stojkoski}}, \bibinfo {author} {\bibfnamefont {T.}~\bibnamefont {Sandev}},
  \bibinfo {author} {\bibfnamefont {L.}~\bibnamefont {Kocarev}},\ and\ \bibinfo
  {author} {\bibfnamefont {A.}~\bibnamefont {Pal}},\ }\href
  {https://doi.org/10.1103/PhysRevE.104.014121} {\bibfield  {journal} {\bibinfo
   {journal} {Phys. Rev. E}\ }\textbf {\bibinfo {volume} {104}},\ \bibinfo
  {pages} {014121} (\bibinfo {year} {2021})}\BibitemShut {NoStop}%
\bibitem [{\citenamefont {Gripenberg}(1983)}]{gripenberg1983stationary}%
  \BibitemOpen
  \bibfield  {author} {\bibinfo {author} {\bibfnamefont {G.}~\bibnamefont
  {Gripenberg}},\ }\href@noop {} {\bibfield  {journal} {\bibinfo  {journal}
  {Journal of Mathematical Biology}\ }\textbf {\bibinfo {volume} {17}},\
  \bibinfo {pages} {371} (\bibinfo {year} {1983})}\BibitemShut {NoStop}%
\bibitem [{\citenamefont {Ben-Ari}\ \emph {et~al.}(2019)\citenamefont
  {Ben-Ari}, \citenamefont {Roitershtein},\ and\ \citenamefont
  {Schinazi}}]{ben2019random}%
  \BibitemOpen
  \bibfield  {author} {\bibinfo {author} {\bibfnamefont {I.}~\bibnamefont
  {Ben-Ari}}, \bibinfo {author} {\bibfnamefont {A.}~\bibnamefont
  {Roitershtein}},\ and\ \bibinfo {author} {\bibfnamefont {R.~B.}\ \bibnamefont
  {Schinazi}},\ }\href@noop {} {\bibfield  {journal} {\bibinfo  {journal}
  {Electronic journal of probability}\ }\textbf {\bibinfo {volume} {24}}
  (\bibinfo {year} {2019})}\BibitemShut {NoStop}%
\bibitem [{\citenamefont {Boyer}\ and\ \citenamefont
  {Romo-Cruz}(2014)}]{boyer_solvable_2014}%
  \BibitemOpen
  \bibfield  {author} {\bibinfo {author} {\bibfnamefont {D.}~\bibnamefont
  {Boyer}}\ and\ \bibinfo {author} {\bibfnamefont {J.~C.~R.}\ \bibnamefont
  {Romo-Cruz}},\ }\href {https://doi.org/10.1103/PhysRevE.90.042136} {\bibfield
   {journal} {\bibinfo  {journal} {Physical Review E}\ }\textbf {\bibinfo
  {volume} {90}},\ \bibinfo {pages} {042136} (\bibinfo {year} {2014})},\
  \bibinfo {note} {publisher: American Physical Society}\BibitemShut {NoStop}%
\bibitem [{\citenamefont {Boyer}\ \emph {et~al.}(2017)\citenamefont {Boyer},
  \citenamefont {Evans},\ and\ \citenamefont {Majumdar}}]{boyer_long_2017}%
  \BibitemOpen
  \bibfield  {author} {\bibinfo {author} {\bibfnamefont {D.}~\bibnamefont
  {Boyer}}, \bibinfo {author} {\bibfnamefont {M.~R.}\ \bibnamefont {Evans}},\
  and\ \bibinfo {author} {\bibfnamefont {S.~N.}\ \bibnamefont {Majumdar}},\
  }\href {https://doi.org/10.1088/1742-5468/aa58b6} {\bibfield  {journal}
  {\bibinfo  {journal} {Journal of Statistical Mechanics: Theory and
  Experiment}\ }\textbf {\bibinfo {volume} {2017}},\ \bibinfo {pages} {023208}
  (\bibinfo {year} {2017})},\ \bibinfo {note} {publisher: IOP
  Publishing}\BibitemShut {NoStop}%
\bibitem [{\citenamefont {Falcón-Cortés}\ \emph {et~al.}(2017)\citenamefont
  {Falcón-Cortés}, \citenamefont {Boyer}, \citenamefont {Giuggioli},\ and\
  \citenamefont {Majumdar}}]{falcon-cortes_localization_2017}%
  \BibitemOpen
  \bibfield  {author} {\bibinfo {author} {\bibfnamefont {A.}~\bibnamefont
  {Falcón-Cortés}}, \bibinfo {author} {\bibfnamefont {D.}~\bibnamefont
  {Boyer}}, \bibinfo {author} {\bibfnamefont {L.}~\bibnamefont {Giuggioli}},\
  and\ \bibinfo {author} {\bibfnamefont {S.~N.}\ \bibnamefont {Majumdar}},\
  }\href {https://doi.org/10.1103/PhysRevLett.119.140603} {\bibfield  {journal}
  {\bibinfo  {journal} {Physical Review Letters}\ }\textbf {\bibinfo {volume}
  {119}},\ \bibinfo {pages} {140603} (\bibinfo {year} {2017})}\BibitemShut
  {NoStop}%
\bibitem [{\citenamefont {Santos}\ and\ \citenamefont
  {F}(2018)}]{santos_non-gaussian_2018}%
  \BibitemOpen
  \bibfield  {author} {\bibinfo {author} {\bibfnamefont {D.}~\bibnamefont
  {Santos}}\ and\ \bibinfo {author} {\bibfnamefont {M.~A.}\ \bibnamefont {F}},\
  }\href {https://doi.org/10.3390/fractalfract2030020} {\bibfield  {journal}
  {\bibinfo  {journal} {Fractal and Fractional}\ }\textbf {\bibinfo {volume}
  {2}},\ \bibinfo {pages} {20} (\bibinfo {year} {2018})},\ \bibinfo {note}
  {number: 3 Publisher: Multidisciplinary Digital Publishing
  Institute}\BibitemShut {NoStop}%
\bibitem [{\citenamefont {Campos}\ and\ \citenamefont
  {Méndez}(2019)}]{campos_recurrence_2019}%
  \BibitemOpen
  \bibfield  {author} {\bibinfo {author} {\bibfnamefont {D.}~\bibnamefont
  {Campos}}\ and\ \bibinfo {author} {\bibfnamefont {V.}~\bibnamefont
  {Méndez}},\ }\href {https://doi.org/10.1103/PhysRevE.99.062137} {\bibfield
  {journal} {\bibinfo  {journal} {Physical Review E}\ }\textbf {\bibinfo
  {volume} {99}},\ \bibinfo {pages} {062137} (\bibinfo {year} {2019})},\
  \bibinfo {note} {publisher: American Physical Society}\BibitemShut {NoStop}%
\bibitem [{\citenamefont {Dahlenburg}\ \emph {et~al.}(2021)\citenamefont
  {Dahlenburg}, \citenamefont {Chechkin}, \citenamefont {Schumer},\ and\
  \citenamefont {Metzler}}]{dahlenburg_stochastic_2021}%
  \BibitemOpen
  \bibfield  {author} {\bibinfo {author} {\bibfnamefont {M.}~\bibnamefont
  {Dahlenburg}}, \bibinfo {author} {\bibfnamefont {A.~V.}\ \bibnamefont
  {Chechkin}}, \bibinfo {author} {\bibfnamefont {R.}~\bibnamefont {Schumer}},\
  and\ \bibinfo {author} {\bibfnamefont {R.}~\bibnamefont {Metzler}},\ }\href
  {https://doi.org/10.1103/PhysRevE.103.052123} {\bibfield  {journal} {\bibinfo
   {journal} {Physical Review E}\ }\textbf {\bibinfo {volume} {103}},\ \bibinfo
  {pages} {052123} (\bibinfo {year} {2021})},\ \bibinfo {note} {publisher:
  American Physical Society}\BibitemShut {NoStop}%
\bibitem [{\citenamefont {Pierce}(2022)}]{pierce_advection-diffusion_2022}%
  \BibitemOpen
  \bibfield  {author} {\bibinfo {author} {\bibfnamefont {J.~K.}\ \bibnamefont
  {Pierce}},\ }\href {http://arxiv.org/abs/2204.07215} {\bibfield  {journal}
  {\bibinfo  {journal} {arXiv:2204.07215 [cond-mat]}\ } (\bibinfo {year}
  {2022})},\ \bibinfo {note} {arXiv: 2204.07215}\BibitemShut {NoStop}%
\bibitem [{\citenamefont {Pierce}(2021)}]{pierce2021stochastic}%
  \BibitemOpen
  \bibfield  {author} {\bibinfo {author} {\bibfnamefont {J.~K.}\ \bibnamefont
  {Pierce}},\ }\emph {\bibinfo {title} {The stochastic movements of individual
  streambed grains}},\ \href@noop {} {Ph.D. thesis},\ \bibinfo  {school}
  {University of British Columbia} (\bibinfo {year} {2021})\BibitemShut
  {NoStop}%
\bibitem [{\citenamefont {Bhat}\ \emph {et~al.}(2016)\citenamefont {Bhat},
  \citenamefont {Bacco},\ and\ \citenamefont {Redner}}]{bhat_stochastic_2016}%
  \BibitemOpen
  \bibfield  {author} {\bibinfo {author} {\bibfnamefont {U.}~\bibnamefont
  {Bhat}}, \bibinfo {author} {\bibfnamefont {C.~D.}\ \bibnamefont {Bacco}},\
  and\ \bibinfo {author} {\bibfnamefont {S.}~\bibnamefont {Redner}},\ }\href
  {https://doi.org/10.1088/1742-5468/2016/08/083401} {\bibfield  {journal}
  {\bibinfo  {journal} {Journal of Statistical Mechanics: Theory and
  Experiment}\ }\textbf {\bibinfo {volume} {2016}},\ \bibinfo {pages} {083401}
  (\bibinfo {year} {2016})},\ \bibinfo {note} {publisher: IOP
  Publishing}\BibitemShut {NoStop}%
\bibitem [{\citenamefont {Eliazar}(2017)}]{eliazar_branching_2017}%
  \BibitemOpen
  \bibfield  {author} {\bibinfo {author} {\bibfnamefont {I.}~\bibnamefont
  {Eliazar}},\ }\href {https://doi.org/10.1209/0295-5075/120/60008} {\bibfield
  {journal} {\bibinfo  {journal} {EPL (Europhysics Letters)}\ }\textbf
  {\bibinfo {volume} {120}},\ \bibinfo {pages} {60008} (\bibinfo {year}
  {2017})},\ \bibinfo {note} {publisher: IOP Publishing}\BibitemShut {NoStop}%
\bibitem [{\citenamefont {Eliazar}\ and\ \citenamefont
  {Reuveni}(2020)}]{eliazar_mean-performance_2020}%
  \BibitemOpen
  \bibfield  {author} {\bibinfo {author} {\bibfnamefont {I.}~\bibnamefont
  {Eliazar}}\ and\ \bibinfo {author} {\bibfnamefont {S.}~\bibnamefont
  {Reuveni}},\ }\href {https://doi.org/10.1088/1751-8121/abae8c} {\bibfield
  {journal} {\bibinfo  {journal} {Journal of Physics A: Mathematical and
  Theoretical}\ }\textbf {\bibinfo {volume} {53}},\ \bibinfo {pages} {405004}
  (\bibinfo {year} {2020})},\ \bibinfo {note} {publisher: IOP
  Publishing}\BibitemShut {NoStop}%
\bibitem [{\citenamefont {Oberhettinger}(2014)}]{oberhettinger_fourier_2014}%
  \BibitemOpen
  \bibfield  {author} {\bibinfo {author} {\bibfnamefont {F.}~\bibnamefont
  {Oberhettinger}},\ }\href@noop {} {\emph {\bibinfo {title} {Fourier
  transforms of distributions and their inverses: a collection of tables}}},\
  Vol.~\bibinfo {volume} {16}\ (\bibinfo  {publisher} {Academic press},\
  \bibinfo {year} {2014})\BibitemShut {NoStop}%
\end{thebibliography}
\end{document}